\documentclass[trackchanges]{aastex63}
\usepackage{gensymb}
\usepackage{threeparttable}

\def\kms       {km~s$^{-1}$}
\def\masy      {mas~yr$^{-1}$}

\def\jybeam    {Jy~beam$^{-1}$}

\def\h       {\ifmmode{^{\rm h}}\else{$^{\rm h}$}\fi}
\def\m       {\ifmmode{^{\rm m}}\else{$^{\rm m}$}\fi}
\def\s       {\ifmmode{^{\rm s}}\else{$^{\rm s}$}\fi}
\def\deg     {\ifmmode{^{\circ}}\else{$^{\circ}$}\fi}
\def\decdeg  {\ifmmode{{\rlap.}^{\circ}} \else ${\rlap.}^{\circ}$\fi}
\def\decs    {\ifmmode{{\rlap.}^{\rm s}} \else ${\rlap.}^{\rm s}$\fi}
\def\decas   {\ifmmode{{\rlap.}{''}}\else{${\rlap.}{''}$}\fi}

\def\VLSR  {V$_{\rm LSR}$}

\def\hho   {H$_2$O}

\def\HIP   {$Hipparcos$}
\def\Gaia  {$Gaia$}
\def\K     {$K$}
\def\MK    {$M_K$}

\def\mK    {$m_K$}

\def\Rs    {$R_{\ast}$}

\def\mux   {$\mu_x$}
\def\muy   {$\mu_y$}

\def\HIP   {$Hipparcos$}

\def\dx    {$\Delta x$}
\def\dy    {$\Delta y$}
\def\ra    {$\alpha_{\rm J2000}$}
\def\dec   {$\delta_{\rm J2000}$}

\def\tsrc  {RR Aql}       

\usepackage{float}

\usepackage{color}
\definecolor{mygreen}{RGB}{44,85,17}                
\definecolor{myblue}{RGB}{34,31,217}                 
\definecolor{mybrown}{RGB}{194,164,113}          
\definecolor{myred}{RGB}{255,66,56}                 
\definecolor{mygrey}{RGB}{211,211,211}                 

\received{December 24, 2021}
\revised{\today}
\accepted{April 22, 2022}
\submitjournal{ApJ}

\shorttitle{A Trigonometric Parallax for RR Aql and the Mira PL Relation}
\shortauthors{Y. Sun et al}

\begin{document}

\title{A VLBA Trigonometric Parallax for RR Aql and the Mira PL Relation}

\correspondingauthor{Bo Zhang}
\email{zb@shao.ac.cn}

\author[0000-0002-8604-5394]{Yan Sun}
\affiliation{Shanghai Astronomical Observatory, Chinese Academy of Sciences\\
80 Nandan Road, Shanghai 200030, China}

\affiliation{University of Chinese Academy of Sciences \\
No.19 (A) Yuquan Rd. Shijingshan, Beijing, 100049, China}

\author[0000-0003-1353-9040]{Bo Zhang}
\affiliation{Shanghai Astronomical Observatory, Chinese Academy of Sciences\\
80 Nandan Road, Shanghai 200030, China}

\author{Mark J. Reid}
\affiliation{Center for Astrophysics~$\vert$~Harvard \& Smithsonian\\
60 Garden Street, Cambridge, MA 02138, USA}

\author{Shuangjing Xu}
\affiliation{Shanghai Astronomical Observatory, Chinese Academy of Sciences\\
80 Nandan Road, Shanghai 200030, China}

\affiliation{Korea Astronomy and Space Science Institute, 776 Daedeok-daero, Yuseong-gu, Daejeon 34055, Republic of Korea}

\author{Shiming Wen}
\affiliation{Shanghai Astronomical Observatory, Chinese Academy of Sciences\\
80 Nandan Road, Shanghai 200030, China}

\author{Jingdong Zhang}
\affiliation{Shanghai Astronomical Observatory, Chinese Academy of Sciences\\
80 Nandan Road, Shanghai 200030, China}

\affiliation{University of Chinese Academy of Sciences \\
No.19 (A) Yuquan Rd. Shijingshan, Beijing, 100049, China}

\author{Xingwu Zheng}
\affiliation{
School of Astronomy and Space Science, Nanjing University\\
22 Hankou Road, Nanjing  210093, China}




\begin{abstract}
We report VLBA observations of 22 GHz \hho\ and 43 GHz SiO masers toward
the Mira variable \tsrc.
By fitting the SiO maser emission to a circular ring, we estimate the
absolute stellar position of \tsrc\ and find agreement with \Gaia\
astrometry to within the joint uncertainty of $\approx1$ mas.
Using the maser astrometry we measure a stellar parallax of 2.44 $\pm$
0.07 mas, corresponding to a distance of 410$^{+12}_{-11}$ pc. The maser
parallax deviates significantly from the \Gaia\ EDR3 parallax of 1.95
$\pm$ 0.11 mas, indicating a $3.8\sigma$ tension between radio and
optical measurements. 
This tension is most likely caused by optical photo-center
variations limiting the Gaia astrometric accuracy for this Mira
variable.
Combining infrared magnitudes with parallaxes for \tsrc\ and other
Miras, we fit a period-luminosity relation using a Bayesian
approach with MCMC sampling and a strong prior for the slope of --3.60
$\pm$ 0.30 from the LMC. We find a $K$-band zero-point (defined at
logP(days) = 2.30) of --6.79 $\pm$ 0.15 mag using VLBI
parallaxes and --7.08 $\pm$ 0.29 mag using \Gaia\ parallaxes.
The \Gaia\ zero-point is statistically consistent with the more accurate
VLBI value.

\end{abstract}


\section{Introduction} \label{sec:intro}

Mira variables in the Large Magellanic Could (LMC) show a well defined
period-luminosity relation (PLR) in the infrared
\citep{1981Natur.291..303G, 2008MNRAS.386..313W,
2017AJ....154..149Y,2021ApJ...919...99I}.
Assuming the PLR slope from the LMC, zero-points of the PLRs for
Miras have also been determined in other galaxies, e.g.,
M~33~\citep{2018AJ....156..112Y}, NGC~4258~\citep{2018ApJ...857...67H}
and NGC 1559~\citep{2020ApJ...889....5H}.
Compared with Cepheids, Miras are brighter at infrared
wavelengths~\citep{2014IAUS..298...40F} and have a larger range of ages
and greater numbers~\citep{2000ASPC..203...71E}. Miras can also provide
distances rivaling those of Cepheids~\citep{2021ApJ...919...99I}.
Thus, Miras will be important in the era of large space infrared
telescopes, e.g., the James Webb Space Telescope (JWST).
%
However, there are some issues regarding the use of the PLR of Miras
which need to be addressed.
First, more observational evidence that the PLR of Miras in different
galaxies is insensitive to metallicity variations is desirable. Second,
if the PLR of Miras is universal, the zero-point for the PLR should be
better determined.
%
Third, using the PLR of LMC Miras to estimate distances to other
galaxies includes a component of systematic error due to the assumed
distance to the LMC;
currently the most accurate distance modulus of $18.477\pm0.026$
mag was published by~\citet{2019Natur.567..200P}.

The \Gaia\ Early Data Release 3 (EDR3) provides parallaxes for more than
1000 Miras with formal uncertainties of $\le$ 0.5 mas. This provides an
opportunity to derive a Galactic PLR for Miras independently of the LMC.
Thus, it is important to check the \Gaia\ parallaxes with independent
methods of comparable or better accuracy, e.g., using Very Long Baseline
Interferometry (VLBI).  As shown in \citet{2019ApJ...875..114X}, there
are significant discrepancies between the parallaxes from \Gaia\ and
VLBI for Miras.  One likely reason for these discrepancies involves
optical photo-center variations, owing to a small number of very large
convective cells~\citep{2011A&A...528A.120C} or hot
spots~\citep{2009ApJ...707..632L}.
Since Miras typically have diameters of $\approx 2$ AU
\citep{1981ASSL...89..353W}, which imply angular sizes of twice
their parallaxes, photo-center variations of, say, 10\% could lead to
astrometric noise of 20\% of a parallax. To further complicate this
problem, Miras typically have periods near one year
~\citep{1981ASSL...89..353W}, and thus photocenter variations
can easily correlate with a yearly parallax signature.

%
In order to determine the PLR for Galactic Miras, accurate parallaxes
for 5 Miras have been obtained with Very Long Baseline Array
(VLBA) observations and 17 have been obtained with the VLBI Exploration
of Radio Astrometry (VERA), as part of a key project started in 2004
\citep{2009asrp.proc...58N,2020PASJ...72...50V}.  Since the total number
of VLBI parallaxes for Miras is modest, adding more is important.
%
%
\tsrc\ is an oxygen rich Mira with a period of 396 days with
strong circumstellar OH, \hho\ and SiO
masers~\citep{1990ApJS...74..911B}.
Although its trigonometric parallax has been estimated optically by
\HIP~\citep{2007A&A...474..653V} and in the radio by observations of its
OH masers~\citep{2007A+A...472..547V}, the uncertainties in these
measurements are too large to be useful as a check on \Gaia.
In this paper, we present VLBI parallax measurements of 22 GHz \hho\ and
43 GHz SiO maser emission toward \tsrc\ using the VLBA.  In
Section~\ref{sec:obs}, we describe the observations and analysis
procedures.  In Section~\ref{sec:results}, we present the positions of
both \hho\ and SiO maser spots in \tsrc\ relative to extragalactic
quasars at different epochs and the derived parallaxes and proper
motions.  In Section~\ref{sec:abspos}, we compare the absolute positions
of \Gaia\ and our VLBI results.
Then in Section~\ref{sec:plr}, we fit a PLR using the VLBI and
\Gaia\ parallaxes separately, using a Bayesian approach with MCMC
sampling which avoids hand editing of data by decreasing the penalty in
the likelihood function for outliers compared to using a Gaussian data
uncertainty.
Finally, we summarize our studies and discuss the future outlook
for the PLR of Miras in Section~\ref{sec:sum}.

\section{Observation and Data reduction} 
\label{sec:obs}

We conducted phase-referenced observations of \tsrc\ (\ra\ =
19\h57\m36\decs0334/36\decs0377, \dec\ = --01\degr 53\arcmin
12\decas157/12\decas203 for \hho/SiO masers, see Table~\ref{tab:source}
for details.) relative to a background quasar J1947--0103~(\ra\ =
19\h47\m43\decs7837, \ra\ = --01\degr03\arcmin 24\decas528) at 22 and 43
GHz under the National Radio Astronomy Observatory's\footnote{The
National Radio Astronomy Observatory is a facility of the National
Science Foundation operated under cooperative agreement by Associated
Universities, Inc.} VLBA program BZ069. These observations included six
epochs spanning about one year from 2017/10/29 to 2018/11/02 at
intervals of roughly two months.
An observation on 2018/08/21 failed with no detections of any quasars or
masers on all of the interferometer baselines; the reason for the failure
is unknown.
Each epoch used a 6-h track which included multiple observing blocks of
25 min of fast-switching between RR Aql and a background quasar
J1947--0103\footnote{Another quasar, J2001--0042, was also observed, but
proved too weak to use.}.  Scan durations for the maser source were 60
and 40 seconds, yielding typical on-source times of 40 and 20 seconds at
22 and 43 GHz, respectively.  For better interferometer uv-coverage at
each band, we alternated the fast-switching blocks at 22 and 43 GHz. The
data were taken in left- and right-circular polarizations with four
adjacent bands of 16 MHz bandwidth, with the maser emission placed in
the third band.

We observed three International Celestial Reference Frame (ICRF)
sources: 3C345, 1638+572 and 1740+521 at the beginning, middle and end
of the observations, in order to calibrate electronic delay and phase
offsets among different bands.
In order to calibrate tropospheric delays, we adopted the methods
described in~\citet{2009ApJ...695..287R}.  We included three 0.5-h
``geodetic blocks'' at the beginning, middle and end of tracks. Geodetic
blocks were recorded in left-circular polarization with eight 16-MHz
bands spanning 480 MHz between 23.5 and 24.0 GHz. The eight bands were
spaced in a ``minimum redundancy configuration'' to uniformly sample
frequency differences.

The data were correlated in two passes using the DiFX software
correlator~\citep{2007PASP..119..318D} in Socorro, NM. One pass
generated 32 spectral channels for all the data, and another pass
generated 1000 spectral channels (in “zoom” mode) only for the single
(dual-polarized) frequency band containing the masers, 
yielding a spectral channels spaced by 8 kHz, corresponding to velocity
resolutions of 0.11 and 0.06 \kms\ for \hho\ and SiO maser emissions,
respectively.

Data calibration and imaging were performed using the Astronomical Image
Processing System (AIPS; \citealt{1996ASPC..101...37V}). For phase
calibration, we corrected for the effects of diurnal feed rotation,
errors in the Earth Orientation Parameters (EOP) and the positions of
the masers and the background source adopted in the correlation. We
removed instrumental delays using the data from 3C345 and calibrated
ionospheric and tropospheric delays using the Global Ionospheric Model,
provided by the International GNSS Service, and data from the geodetic
blocks~\citep{2014ARA&A..52..339R}.
For amplitude calibration, we adopted the VLBA's recommended system
equivalent flux densities (SEFD) for each antenna and feed, using gain
curves and system temperatures.

\begin{figure}[ht!]
    \centering
    \includegraphics[angle=0,scale=0.75]{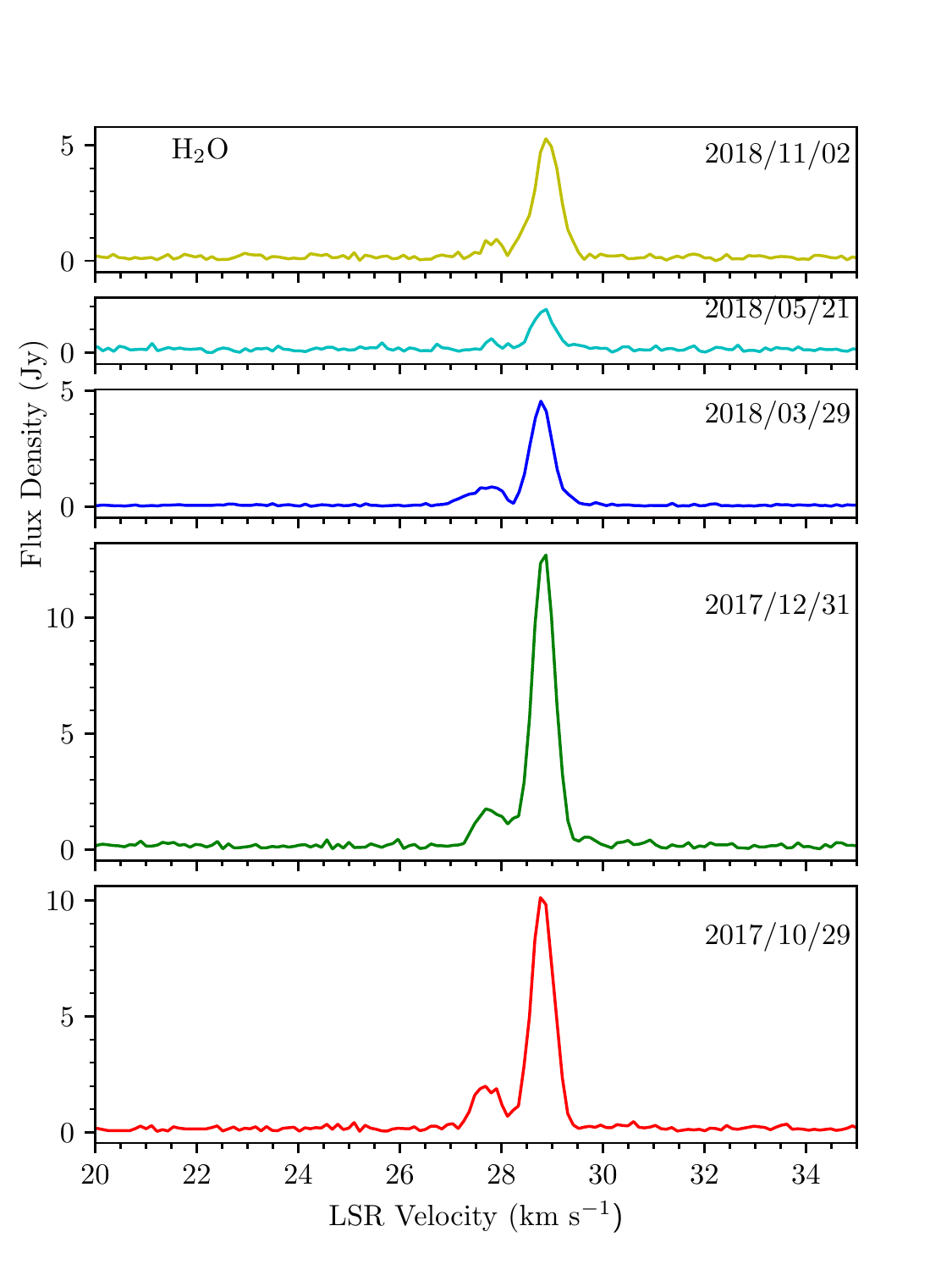}
    \includegraphics[angle=0,scale=0.75]{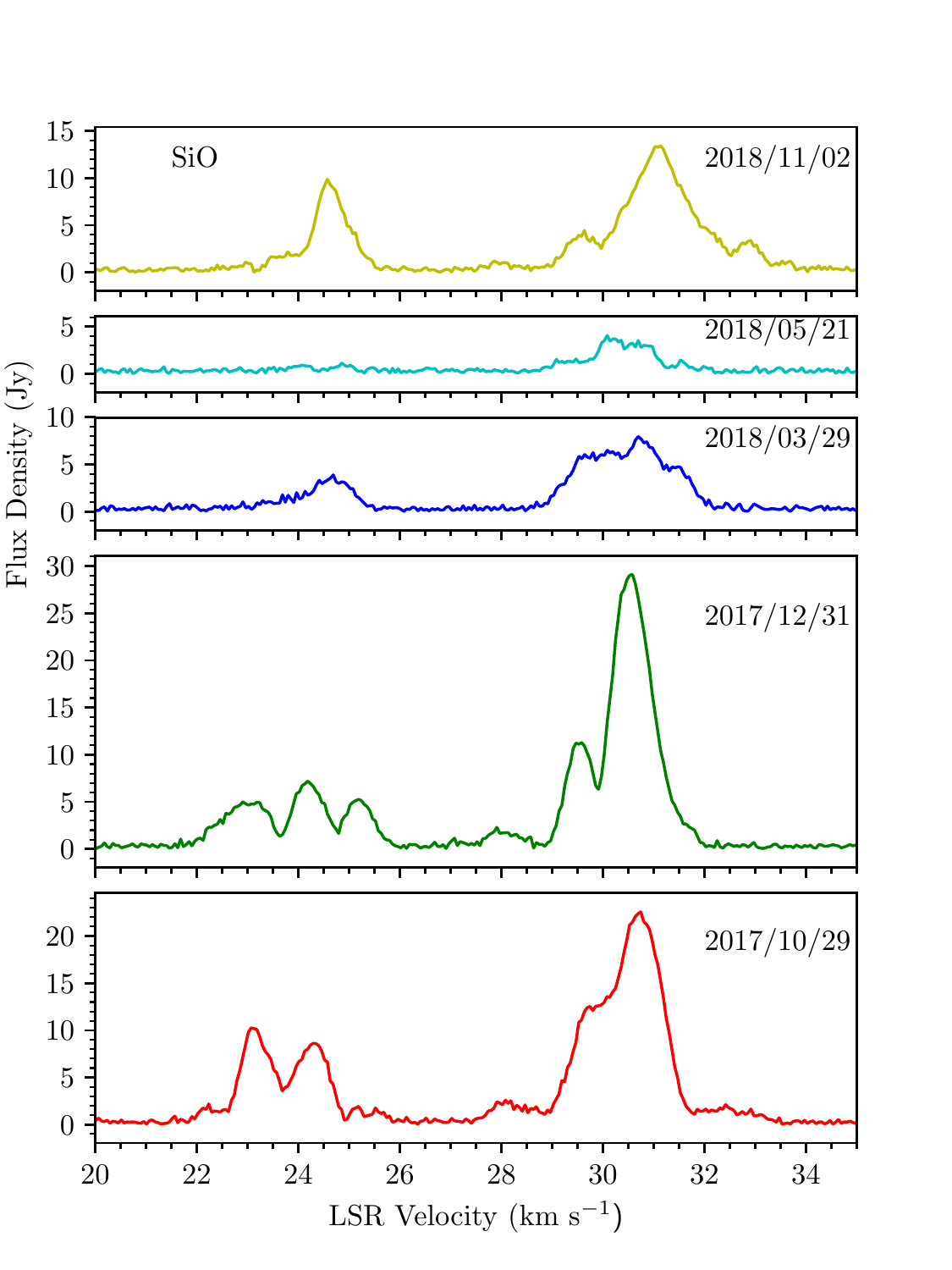}
    \caption{Vector-averged interferometer spectra of \hho\ ({\it left panel})
    and SiO ({\it right panel}) maser emission toward RR Aql from the Kitt Peak
    to Los Alamos baseline at five epochs.
    \label{fig:spec}}
\end{figure}
As shown in Figure~\ref{fig:spec}, we found  strong \hho\ and SiO maser
emissions with peaks at the velocity of Local Standard Rest~(\VLSR)
values of around 29 and 31 \kms\ for all epochs.
In this paper, we measure parallax using the maser channels
centered at 28.8 and 28.9 \kms\ for \hho\ masers and 30.9 \kms\ for the
SiO masers.
Since the water maser emissions were heavily resolved on the longer
baselines, we phase referenced those data using the quasar J1947--0103
and only used the data from the VLBA inner-five antennas: Fort
Davis~(FD), Kitt Peak~(KP), Los Alamos~(LA), Owens Valley~(OV) and Pie
Town~(PT). These antennas are located close to the center of the
array, with relatively short baselines.
However, on 2017/10/29, there were clock jumps at the PT antenna, which
were evident in plots of interferometric delay versus time using the
strongest calibrator, and we discarded the data involving PT.  In
addition, the PT antenna was unavailable for the observation on
2017/12/31 owing to a maintenance issue.
For the SiO masers, we used the 30.9 \kms\ maser spot as the
interferometer phase reference (i.e., ``inverse'' phase-referencing).

\begin{figure*}
   \fig{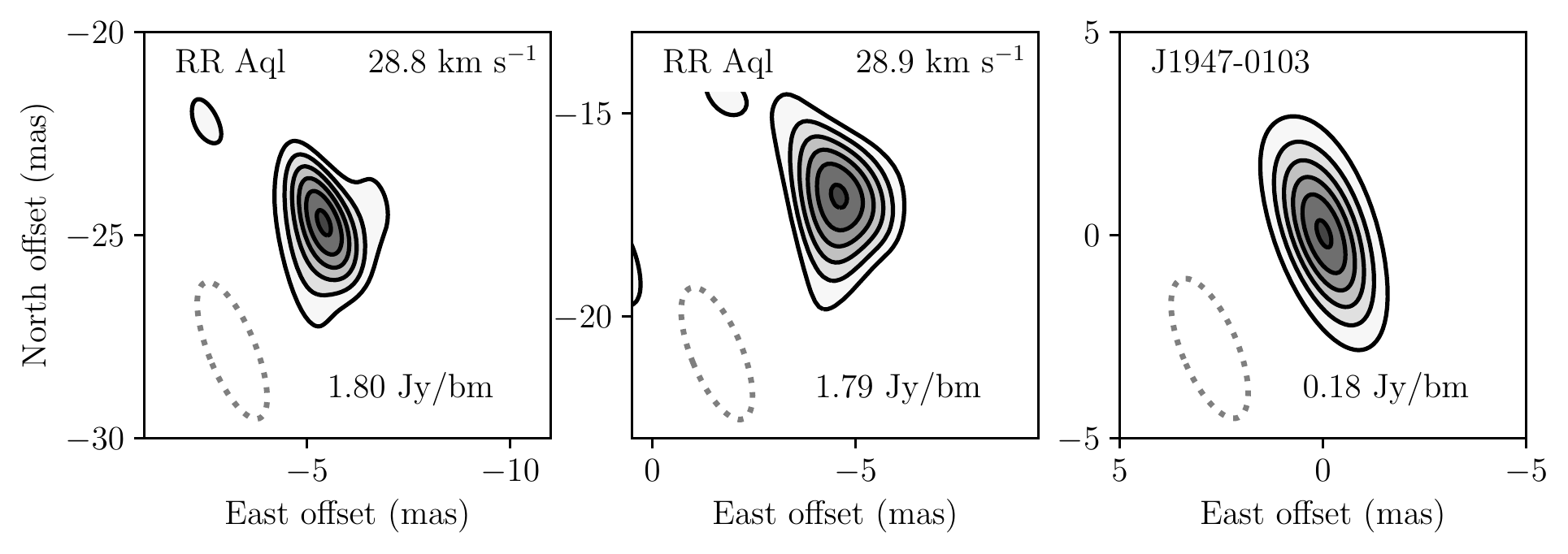}{0.95\textwidth}{}
   \caption{
   Images of the \hho\ maser spots at \VLSR=28.8 \kms ({\it left panel})
   and \VLSR=28.9 \kms\ ({\it middle panel}) toward RR Aql and the
   background  radio source J1947--0103 ({\it right panel}) on 2018/03/29
   using the inner-five antennas of the VLBA.  The restoring beams are shown in
   the lower left corner of each panel. Contour levels are spaced linearly by
   0.25 Jy beam$^{-1}$ for RR Aql and 0.030 Jy beam$^{-1}$ for J1947--0103.}
    \label{fig:K_image}
\end{figure*}

After calibration, we combined the four dual-polarized bands and imaged the
continuum emission of the background source using the AIPS task {\it
IMAGR},
which Fourier transforms the visibility data to make a brightness
image.
The restoring beam for each epoch was calculated using uniform
weighting.  In individual spectral channels, we estimated peak
brightnesses, positions and angular sizes using the task {\it JMFIT},
which fits Gaussian brightness distributions to a portion of an
image.
Table \ref{tab:source} lists basic information for the observed sources,
and Table~\ref{tab:data} gives the measured residual position offsets
for the maser spots relative to the quasar J1947--0103 used for parallax
fitting.  Figure~\ref{fig:K_image} shows an example image of the \hho\
maser spots at \VLSR\ = 28.8 \kms\ and 28.9 \kms\ toward \tsrc\
and the background source J1947--0103.

\begin{deluxetable*}{cccc cccc}
    \tabletypesize{\scriptsize}
    \tablecaption{Positions and Brightnesses}
    \label{tab:source}
    \tablehead{
    \colhead{Source}         & \colhead{R.A. (J2000) (K / Q)} & \colhead{Dec. (J2000)(K / Q) }        & \colhead{$S_p$~(K / Q)} &
    \colhead{$\theta_{sep}$} & \colhead{P.A.}                 & \colhead{\VLSR~(K / Q)}               & \colhead{Beam~(K / Q)} \\
    \colhead{Name}           & \colhead{(h~~~m~~~s)}          & \colhead{(\degr~~~\arcmin~~~\arcsec)} & \colhead{(Jy/beam)} &
    \colhead{(\degree)}      & \colhead{(\degree)}            & \colhead{(km s$^{-1}$)}               & \colhead{(mas~~mas~~\degr)}
    }
    \decimalcolnumbers
    \startdata
    RR Aql      & 19 57 36.0334 /  36.0377  & --01 53 12.157 / 12.203  & 6 / 13           &  ...    & ...     & 28.9 / 30.9  & 3.49$\times$1.3  @ $+$22 / 1.85$\times$0.6  @ $+$23\\
    J1947--0103 & 19 47 43.7837             & --01 03 24.528           & 0.167 / 0.042    & 2.60    &  +159   & ...          & 3.48$\times$1.2  @ $+$22 / 1.74$\times$0.6  @ $+$23\\
    \enddata
    \tablecomments{
    K and Q denote the observing bands, $\sim22$ GHz for the
    \hho\ maser and $\sim43$ GHz for the SiO maser, respectively.
    The second and third columns list the absolute positions of the
    \hho\ maser spot at \VLSR\ = 28.9 \kms\ and of the SiO maser spot
    at \VLSR\ = 30.9 \kms\ on 2018/01/09.
    The fourth and seventh columns give the peak brightnesses ($S_p$)
    and \VLSR\ of the maser spots.  The fifth and sixth columns give the
    separations ($\theta_{sep})$ and position angles (P.A.) east of
    north of the quasar relative to the maser.  The last column gives
    the FWHM size and P.A. of the Gaussian restoring beam at the first
    epoch.  The quasar position is from \url{http://astrogeo.org}. }
\end{deluxetable*}

\begin{deluxetable*}{cccrr}
    \tablecaption{Residual position differences between maser spots and 
    J1947--0103 used for the parallax fits.
    \label{tab:data}
    }
    \tablehead{
     \colhead{Epoch}        & \colhead{Maser}   & \colhead{\VLSR} & \colhead{East offset} & \colhead{North offset} \\
     \colhead{(yyyy/mm/dd)} & \colhead{Species} & \colhead{(\kms)}& \colhead{(mas)}    & \colhead{(mas)}
    }
    \decimalcolnumbers
    \startdata
      2017/10/29 & \hho & 28.9  &   0.043$\pm$0.080   &    -0.051$\pm$0.133  \\
      2017/12/31 &      &       & --2.242$\pm$0.060   &   --8.407$\pm$0.102  \\
      2018/03/29 &      &       & --4.543$\pm$0.129   &  --17.096$\pm$0.180  \\
      2018/05/21 &      &       & --8.113$\pm$0.111   &  --24.495$\pm$0.194  \\
      2018/11/02 &      &       & --22.480$\pm$0.186  &  --46.289$\pm$0.238  \\
 \hline
      2017/10/29 & \hho & 28.8  & --1.067$\pm$0.061   &   --6.565$\pm$0.113  \\
      2017/12/31 &      &       & --3.112$\pm$0.052   &  --15.165$\pm$0.094  \\
      2018/03/29 &      &       & --5.850$\pm$0.138   &  --24.830$\pm$0.215  \\
      2018/05/21 &      &       & --9.058$\pm$0.131   &  --31.472$\pm$0.238  \\
      2018/11/02 &      &       &--23.535$\pm$0.199   &  --53.031$\pm$0.242  \\
 \hline
      2017/10/29 & SiO  & 30.9  & 23.945$\pm$0.038    &    50.273$\pm$0.069  \\
      2017/12/31 &      &       & 21.800$\pm$0.054    &    41.297$\pm$0.097  \\
      2018/03/29 &      &       & 19.286$\pm$0.052    &    30.119$\pm$0.107  \\
      2018/05/21 &      &       & 15.468$\pm$0.132    &    26.109$\pm$0.243  \\
      2018/11/02 &      &       & 1.125 $\pm$0.069    &     0.736$\pm$0.153  \\
    \enddata
    \tablecomments{
     The first column lists the observing epochs. The second and third
     column list the maser species and their \VLSR.  The fourth and
     fifth columns give position offsets relative to \ra\ =
     19\h57\m36\decs0337, \dec\ = --01\degr53\arcmin12\decas148 for the
     \hho\ masers on 2017/10/29, and \ra\ = 19\h57\m36\decs0364,
     \dec\ = --01\degr53\arcmin12\decas242 for the SiO maser on
     2018/11/02. Uncertainties given are statistical only, based
     on thermal noise in the images, and do not include systematic
     effects such as uncompensated atmospheric delays.
    }
\end{deluxetable*}

\section{Results}
\label{sec:results}

\subsection{Maser spots identification for astrometry}
\label{subsec:select}

\subsubsection{Water masers}

\begin{figure*}[ht!]
    \centering
    \includegraphics[angle=0,scale=0.5]{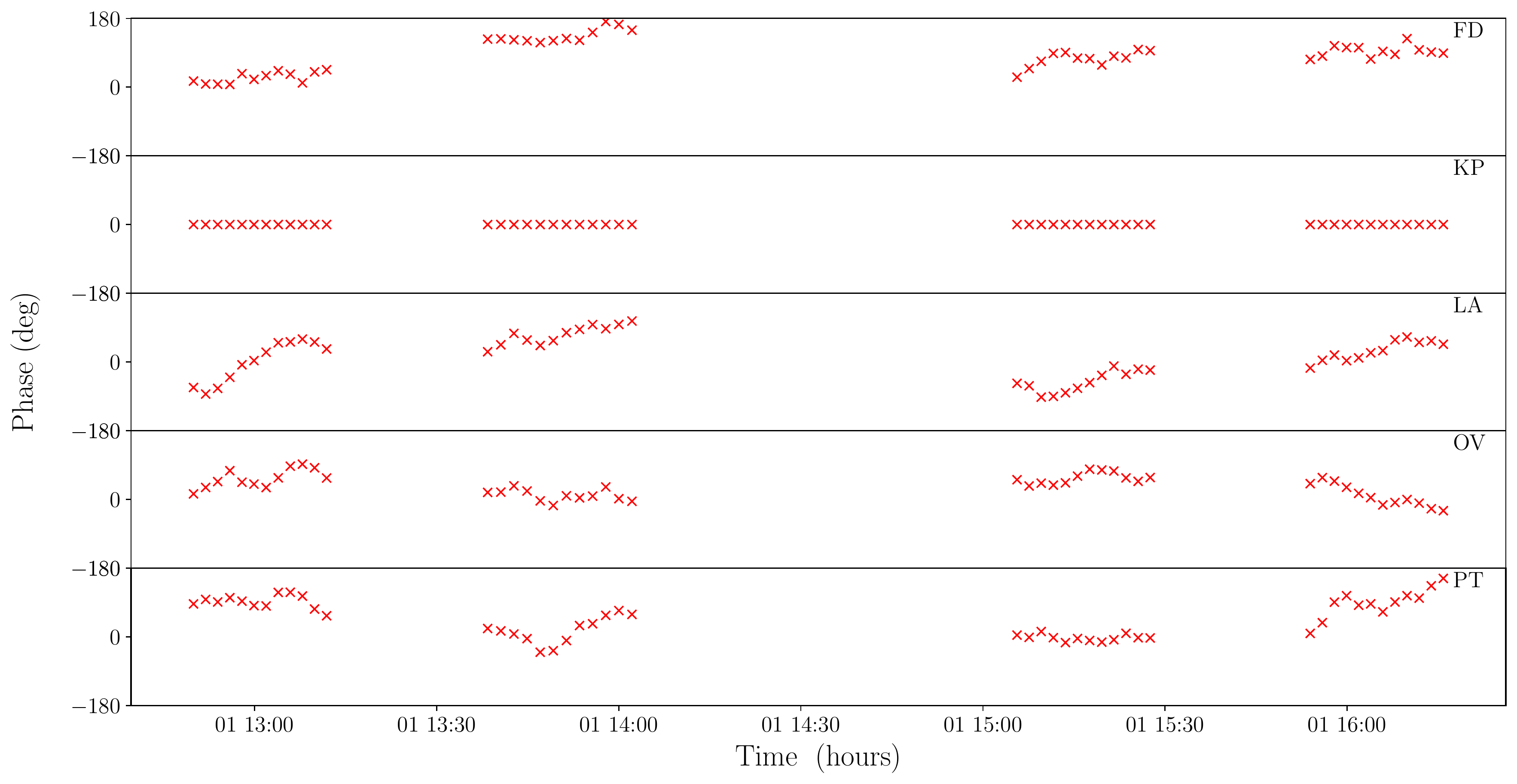}
    \caption{Visibility phase versus observing time of calibrator
    J1947--0103 at 22 GHz on 2018/03/29. 
    The two-letter station codes of the VLBA inner-five antennas are
    labeled on the top-right corner in each panel. FD, KP, LA, OV and PT
    indicate Fort Davis, Kitt Peak, Los Alamos, Owens Valley and Pie
    Town, respectively.
    The reference station was KP. The interval
    between adjacent data points is 2 min.}
    \label{fig:phase}
\end{figure*}

As shown in Figure~\ref{fig:phase}, the phase variations of the
reference source J1947--0103 on 2018/03/29, as an example, are
relatively slow; thus the phases on the baselines of inner-five VLBA
antennas can be confidently connected without 2$\pi$-ambiguities between
adjacent scans separated by 2 min.

\begin{figure*}[ht!]
   \centering
    \includegraphics[angle=0,scale=1.0]{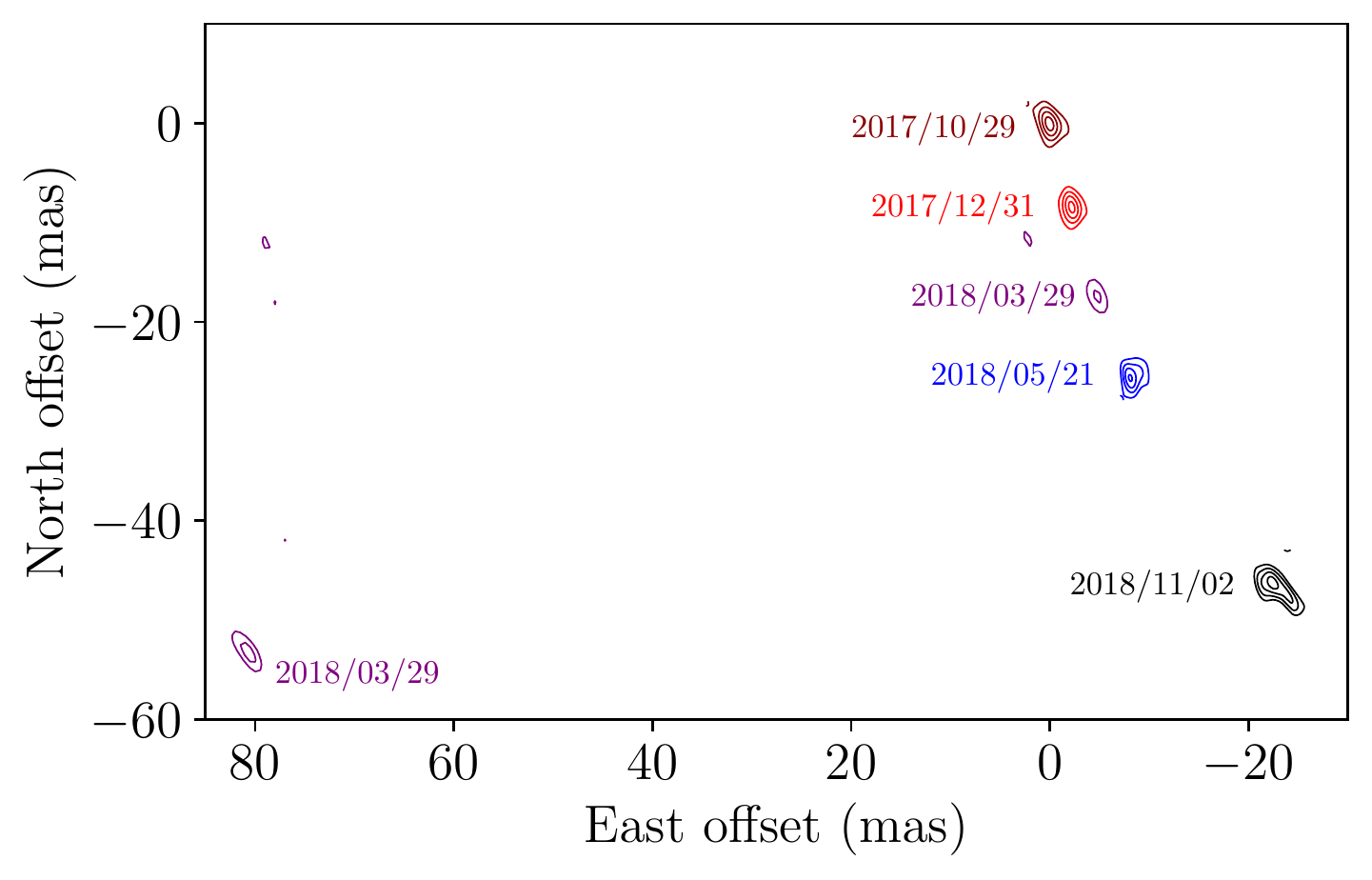}
    \caption{
    Spatial distribution of the \hho\ maser emission at \VLSR\ = 28.9
    \kms\ toward \tsrc\ from VLBA observations at five epochs.  The
    observing epoch is labeled next to each maser spot.  Contour
    levels are integer multiples of  10\% of the peak brightness.
    Note that on 2018/03/29 two maser features are detected.
    \label{fig:h2o_sky}}
\end{figure*}

Water maser emission was detected at \VLSR\ of 27.9 \kms\ from
2017/10/29 to 2018/03/29.  However, by 2018/05/21 its brightness fell
below 0.4 \jybeam\ and was not detected.  This is not unusual for maser
emission from Miras, which also vary over the stellar period.
For parallax measurement, we focus on analyzing the stronger water maser
spots at \VLSR\ of 28.8 and 28.9 \kms, which were clearly detected at all
epochs.
Figure~\ref{fig:h2o_sky} shows the evolution of the 28.9 \kms\ maser
spot relative to the calibrator J1947--0103.  On 2018/03/29 we
detected strong emission at (E,N) of $\approx(-5,-20)$ mas~(left panel)
and weaker emission at $\approx(80,-55)$ mas~(right panel).  Clearly the
weaker spot is unrelated to those tracking smoothly toward the
south-south-west over the five epochs. Therefore, we excluded the weaker
spot from our parallax fitting.  On 2018/11/02 the maser
emission displays a head-tail like structure. This is probably due to
blending two maser spots separated by several mas.  Parallax fitting
with one or the other spot position clearly shows that the stronger
``head'' spot is the continuation of the previous four epochs.

\subsubsection{SiO masers}

\begin{figure*}
    \includegraphics[angle=0,scale=0.7]{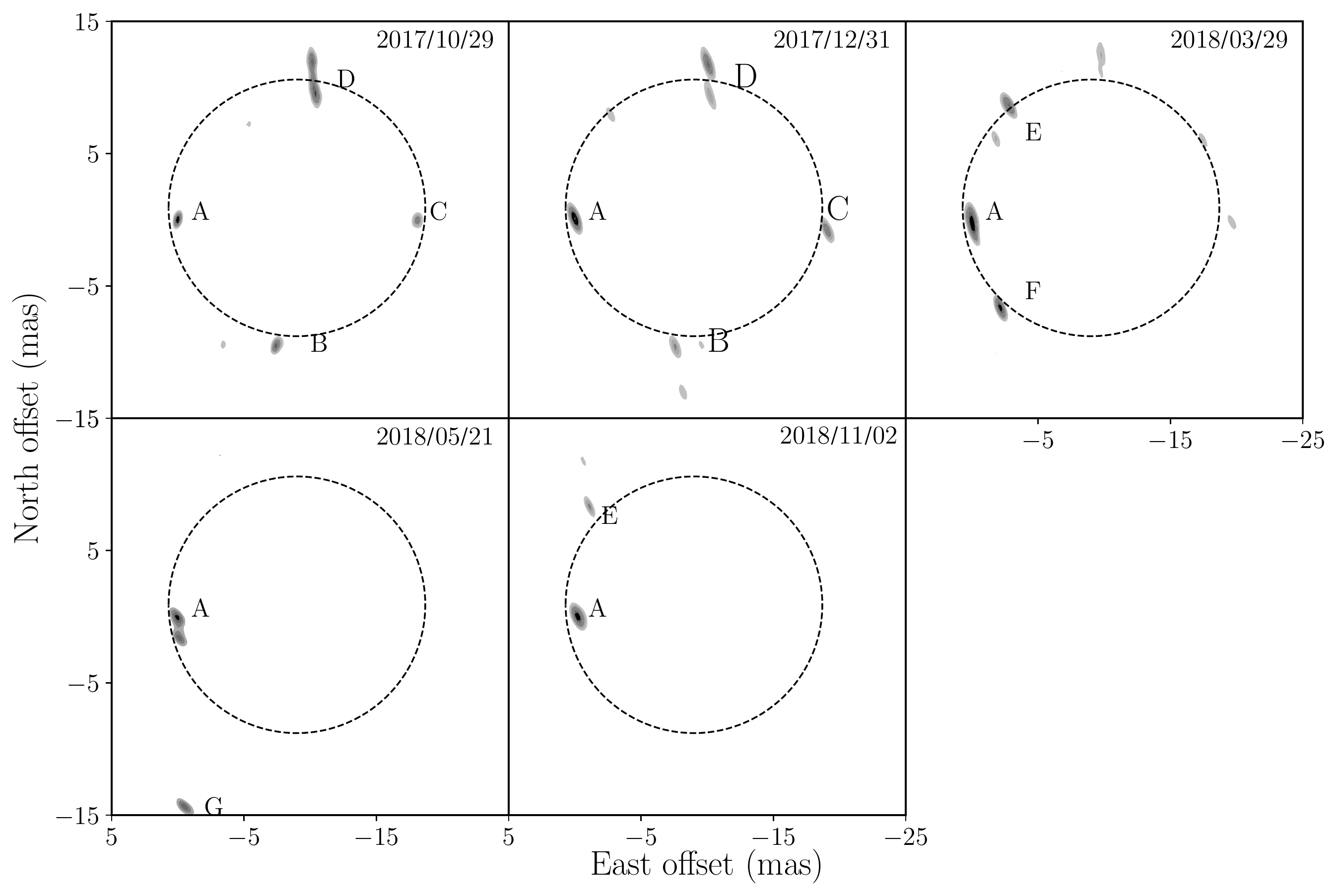}
    \caption{
    Spatial distributions of SiO maser spots toward \tsrc\ from VLBA
    observations at five epochs.  Maser features are designated with letters.  
    The long-lived feature A was used for parallax estimation.
    The dashed circles denote rings fitted to the SiO maser emissions.
    }
    \label{fig:sio_ring}
\end{figure*}

As shown in the right panel of Figure~\ref{fig:spec}, the SiO masers
have more features than the \hho\ masers: there are at least a handful
of SiO features at \VLSR\ ranging from 23 to 31 \kms.  After
phase-referencing to the maser spot at \VLSR\ of 30.9 \kms, we made the
velocity integrated images shown in Figure~\ref{fig:sio_ring}.  There
are clear indications of a clumpy ring-like structure at the first three
epochs; in later epochs the western part of the ring weakens and
vanishes.  Fortunately, maser feature A was detected at all epochs, and
we used this feature to determine the parallax of \tsrc.

\begin{figure}
    \centering
    \includegraphics[angle=0,scale=0.8]{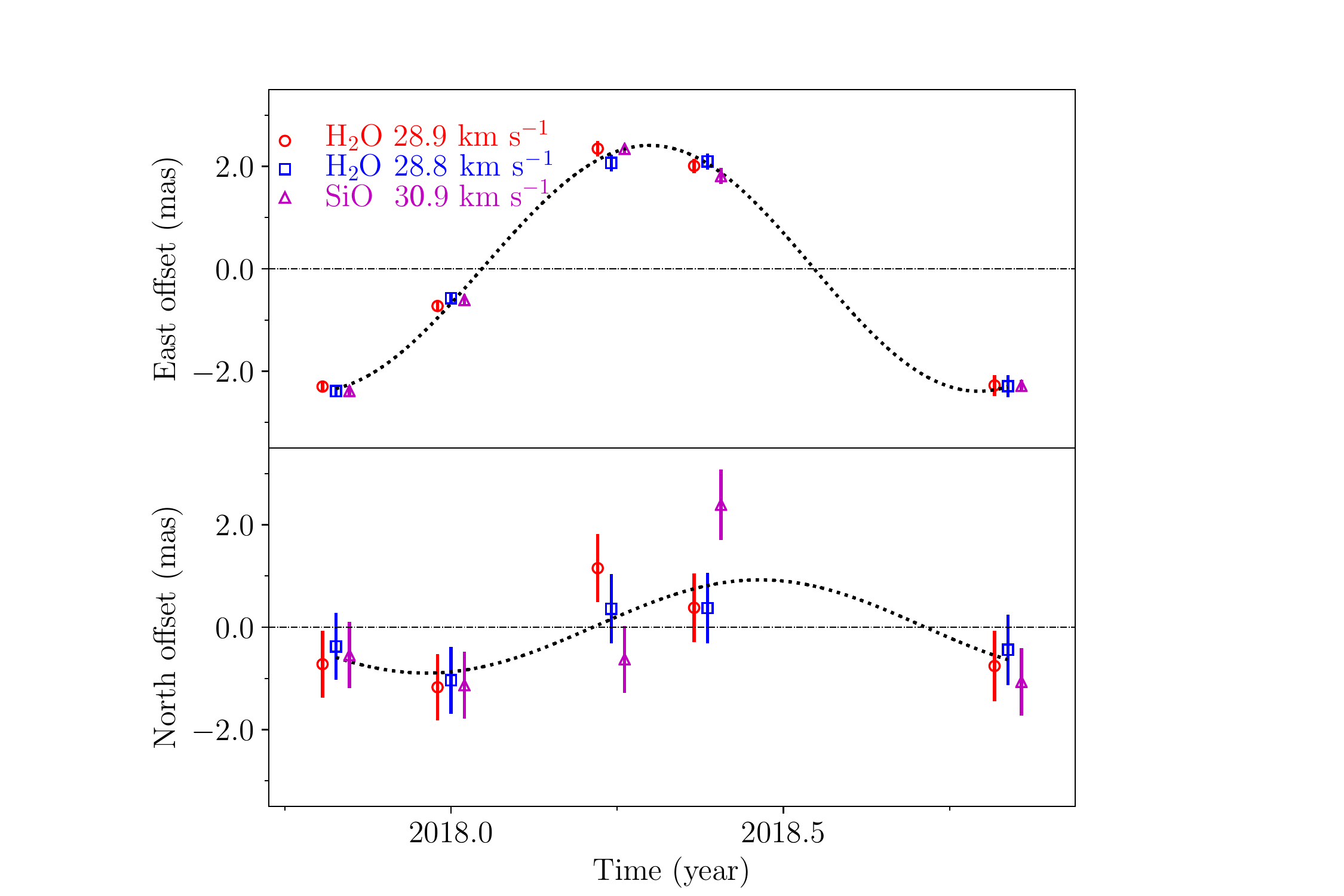}
    \caption{
    East ({\it upper panel}) and north ({\it lower panel}) offsets
    relative to J1907--0103 versus time used for parallax fitting for two \hho\
    maser spots {\it (blue squares and red circles)} and one SiO maser spot 
    {\it (purple triangle)}, with individual constant offset and proper motion
    removed. The \VLSR\ of each maser spot is labeled on the left-top corner in
    the upper panel. The best-fitting model ({\it dotted lines}) denote the
    common parallax curve. Positions are slightly offset in time for clarity.}

    \label{fig:parafit}
\end{figure}

\begin{deluxetable}{ccrr rrrr cc}
    \tabletypesize{\footnotesize} 
    \tablecaption{Parallax and proper motion for \tsrc~\label{tab:para_pm}}
    \tablewidth{0pt}
    \tablehead{
        \colhead{Method} & \colhead{\VLSR} & \colhead{Parallax} & \colhead{\mux} & \colhead{\muy} & \colhead{\dx} & \colhead{\dy} & \colhead{$\chi^2_{\nu}$} & \colhead{$\sigma_x$} & \colhead{$\sigma_y$} \\
        \colhead{}       & \colhead{(\kms)} & \colhead{(mas)}    &
        \colhead{(\masy)} & \colhead{(\masy)} & \colhead{(mas)} &
        \colhead{(mas)} & \colhead{}               & \colhead{(mas)}
        & \colhead{(mas)}
    }
    \startdata
    \hho &    28.9   & 2.44 $\pm$ 0.07  & --22.29 $\pm$ 0.17  &--45.70 $\pm$ 0.85 &--8.94 $\pm$ 0.05 & --22.43 $\pm$ 0.30 & 0.83 & 0.00 & 0.70\\
    \hho &    28.8   & 2.41 $\pm$ 0.09  & --22.29 $\pm$ 0.23  &--45.91 $\pm$ 0.46 &--9.96 $\pm$ 0.08 & --29.39 $\pm$ 0.17 & 0.96 & 0.09 & 0.31\\
    SiO  &    30.9   & 2.44 $\pm$ 0.08  & --22.68 $\pm$ 0.18  &--48.47 $\pm$ 1.46 & 14.86 $\pm$ 0.07 &   26.33 $\pm$ 0.52 & 0.98 & 0.13 & 1.14\\
    Combined   &     & 2.44 $\pm$ 0.07  &                     &                   &                  &                    & 0.98 & 0.00 & 0.00\\
    \Gaia~EDR3 &     & 1.95 $\pm$ 0.11  & --20.42 $\pm$ 0.12  &--47.69 $\pm$ 0.08 &                  &                    &      &      &     \\
    \enddata
    \tablecomments{
        Absolute proper motions are defined as \mux~= $\mu_{\alpha}
        \cos{\delta}$ and \muy~= $\mu_{\delta}$.  Position offsets \dx\
        and \dy\ are for 2018/05/01 (the midpoint between the
        first and last epochs) and are relative to the positions listed
        in Table~\ref{tab:source}.
        $\chi^2_{\nu}$ is the reduced $\chi^2$ of post-fit residuals,
        $\sigma_x$ and $\sigma_y$ are error floors in $x$ and $y$,
        respectively.  `\hho' and `SiO' denote the maser species;
        `Combined' means the parameters derived by combing the data of
        the two \hho\ and one SiO maser spots.  The last row lists the
        parameters from \Gaia\ EDR3.
        }
    \label{tab:parafit}
\end{deluxetable}

\subsection{Parallax and Proper Motion}
\label{subsec:para}

As described in Section~\ref{subsec:select}, we used two \hho\
and one SiO maser spot detected in all five epochs spanning one year.
The positions of maser spots relative to the background source
were modeled by the parallax sinusoid in both coordinates (determined by
one parameter) and a linear proper motion in each coordinate,
and best parameter values were obtained by variance-weighted
least-squares fitting.
The formal position uncertainties listed in Table~\ref{tab:data}
reflect thermal (random) noise in the interferometric images and do not
include systematic uncertainties.  To estimate and allow for
systematic position errors caused by residual atmospheric delays and
variable structure of the maser spots and/or the background quasar, we
added ``systematic errors'' (error floors) for each coordinate in
quadrature with the formal measurement errors and adjusted these to
reach a reduced $\chi^{2}$ for the post-fit residuals near unity for
each coordinate.

Figure~\ref{fig:parafit} shows the positions of the two \hho\
spots and one SiO spot relative to the background source J1947--0103,
with superposed curves showing the best fit parallax sinusoid.
Note that the error bars in the bottom panel are larger than
those in the top panel, indicating the ``error floors'' for the
north-south offsets were larger than those for the east-west offsets.
This occurs because the interferometer beams are larger in the
north-south direction and uncompensated atmospheric delays are more
correlated with them.
Table~\ref{tab:parafit} lists the individual fits (for each maser spot)
and a combined fit (using all three spots), allowing for different
offsets and proper motions among them.

For the combined fits using both \hho\ and SiO masers, to avoid a single
error floor down-weighting the better \hho\ maser data, we combined
measurement uncertainties with the  error floors used in the individual
fits (multiplied by $0.8$) in revised data files, in order to get a
reduced chi-square near unity.
The parallax fitting results are in good agreement within their joint
measurement uncertainties.
Considering that the dominant error is from uncompensated atmospheric
delays, which are essentially the same for all spots and both maser
species, we conservatively inflated the formal parallax uncertainty by
$\sqrt{3}$ (for the 3 spots combined).  We adopt the parallax from the
combined fits, which is 2.44 $\pm$ 0.07 mas, corresponding to a distance
of 410$^{+12}_{-11}$ pc. 
We find that the maser parallax deviates significantly from the
\Gaia\ EDR3 parallax of 1.95 $\pm$ 0.11 mas, indicating a $3.8\sigma$
tension between radio and optical astrometry.  We discuss possible
reasons for this discrepancy in the Section~\ref{sec:abspos}.

\section{Stellar Position of RR Aql}
\label{sec:abspos}

\subsection{Relative positions of masers respect to the central star of
\tsrc}

The absolute stellar positions of Miras determined from VLBI
observations can be used to check the optical \Gaia\ positions,
which might be corrupted by non-uniform circumstellar dust
extinction and photo-center variation. Additionally, knowing the
position of masers relative to the central stars can be crucial to
understanding gas kinematics in the circumstellar material.

As shown in Figure~\ref{fig:sio_ring}, the sky distribution of SiO
masers from our VLBA observations at the first three epochs show a
clumpy ring-like structure.  This has previously been noted for \hho\
maser emission (e.g., from W Hya by~\citealt{1991ASPC...16..375R}) and
SiO maser emission (e.g., from TX Cam by~\citealt{1997ApJ...481L.111K}).
We fit the central position (assumed to be the stellar position)
and the radius of a ring to the maser positions using a least-square
method described by~\citet{2018NatCo...9.2534Y}, which assumes the
distribution of SiO masers lie at a common radius from the central
star. The fitted results are shown in Table~\ref{tab:center}.

\begin{figure*}[ht!]
    \centering
    \includegraphics[angle=0,scale=0.7]{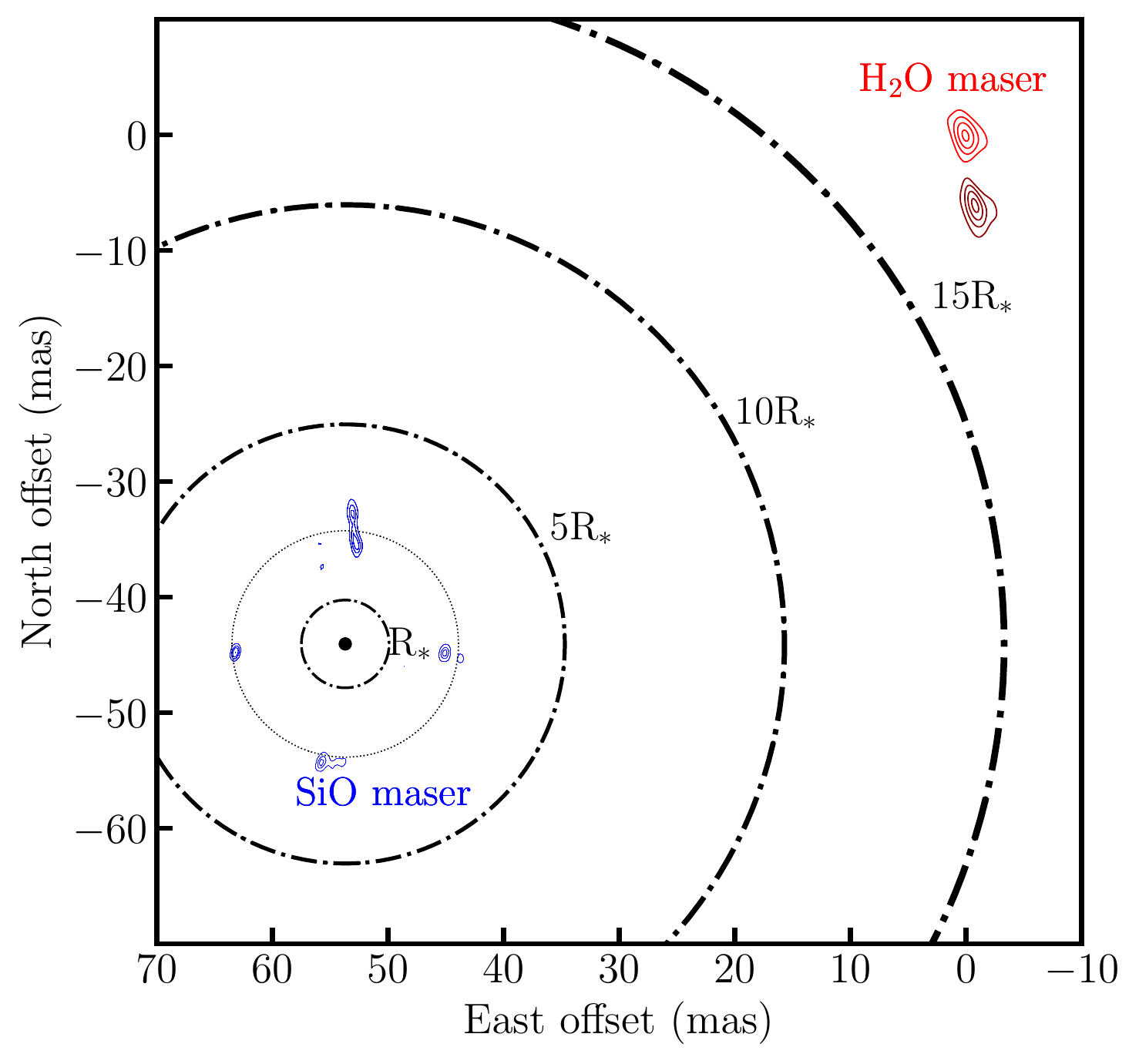}
    \caption{
    The sky distribution SiO and \hho\ masers with respect to the
    central star of \tsrc\ on 2017/10/29.  The dotted line represents
    the fitted SiO maser ring. The circles in dash-dotted lines indicate
    multiples of five stellar radii.
    \label{fig:h2oandsio}}
\end{figure*}

Figure~\ref{fig:h2oandsio} shows relative positions of masers respect to
the central star on 2017/10/29.
Adopting a stellar radius (\Rs) of 3.8 $\pm$ 0.3 mas for \tsrc\ from
multi-epoch mid-infrared interferometric observations by
\citet{2011A&A...532A.134K}, we conclude that the SiO masers are
distributed near 2\Rs,
while the \hho\ maser emission at \VLSR = 28.8 \kms\ and 28.9
\kms\ is located $\approx18$\Rs\ north-west of the central star.
This is consistent with previously reported circumstellar maser
distributions by \citet{1997ApJ...476..327R} and
\citet{2010evn..confE...5R}.

\begin{deluxetable*}{c rr rr r}
    \tablecaption{Center position and radius of SiO maser ring}
    \tablehead{
    \colhead{Epoch}        & \colhead{\ra} & \colhead{$\sigma_\alpha$} & \colhead{\dec} & \colhead{$\sigma_\delta$} & \colhead{Radius} \\
    \colhead{(yyyy/mm/dd)} & \colhead{(h~~~m~~~s)} & \colhead{(mas)} & \colhead{(\degr~~~\arcmin~~~\arcsec)}  & \colhead{(mas)}& \colhead{(mas)}
    }
    \decimalcolnumbers
    \startdata
    2017/10/29  &  19 57 36.03728 & $\pm$0.3   & --01 53 12.1908 & $\pm$1.2  &10.0$\pm$0.1 \\
    2017/12/31  &  19 57 36.03715 & $\pm$0.3   & --01 53 12.1997 & $\pm$1.2  & 9.9$\pm$0.1 \\
    2018/03/29  &  19 57 36.03797 & $\pm$0.3   & --01 53 12.2100 & $\pm$1.2  &10.2$\pm$0.2 \\
    \enddata
\tablecomments{
Columns 2 though 5 list the absolute positions and uncertainties in
Right Ascension and Declination of the center of the SiO maser ring.
Column 6 list the best-fit radii of SiO maser rings.
}
\label{tab:center}
\end{deluxetable*}

\subsection{Comparison of the absolute position of \tsrc\ from VLBA and \Gaia }
\label{ssec:abs_pos}

In order to evaluate differences between optical photo-centers and radio
measurements, we compare the absolute positions of the center of the SiO
ring with stellar positions from \Gaia\ (see Table~\ref{tab:center}).
At the first three epochs the SiO masers have a well-constrained ring
structure, which give very accurate positions for the central star.
Table~\ref{tab:curve} shows the differences between our radio and the
\Gaia\ optical stellar positions (using \Gaia's five astrometric
parameters listed in EDR3); joint uncertainties are estimated by adding
the individual values in quadrature.  There are no statistically
significant differences in either the Right Ascension or Declination
components.  We note that uncertainty in the optical to radio celestial
frame-link also contributes to the position differences; there is
approximately a 1 mas difference in the orientation of radio and optical
frame when measured at brighter magnitudes between the \Gaia\ DR2 and
VLBI data of 26 radio stars \citep{2020A&A...637C...5L}.
%

The \Gaia\ EDR3 ``astrometric excess noise'' value, which measure the
disagreement between the observations and the best-fitting astrometric
model, is 0.84 mas for \tsrc.  As pointed out
by~\citet{2021A&A...649A...2L}, this noise not only includes the effect
of the photo-center variability, but can also come from  modelling
errors (e.g., from excess attitude noise of \Gaia\ satellite).  However,
assuming modelling errors are small, it is likely that photo-center
variability would be near the astrometric excess noise value, which is
$\approx$ 15\% of its angular diameter.  This is possible, since as
shown in the optical interferometric image of the Mira variable $\chi$
Cyg by~\citet{2009ApJ...707..632L}, a single hot spot can offset to the
stellar center by up to 30\% of its diameter.

\begin{deluxetable*}{crrr}
\tablecaption{Differences in positions of \tsrc\ from \Gaia\ EDR3 and VLBA
\label{tab:curve}
}
\tablehead{
\colhead{Epoch} & \colhead{R.A. Difference} & \colhead{Decl. Difference} & \colhead{2D Difference} \\
\colhead{(yyyy/mm/dd)} & \colhead{(mas)} & \colhead{(mas)} & \colhead{(mas)}
}
\decimalcolnumbers
\startdata
  2017/10/29  &   --0.22 $\pm$ 0.41      &  0.59 $\pm$ 1.19 &  0.63$\pm$1.26\\
  2017/12/31  &     0.58 $\pm$ 0.42      &  1.00 $\pm$ 1.20 &  1.16$\pm$1.27\\
  2018/03/29  &   --0.49 $\pm$ 0.44      &  0.64 $\pm$ 1.20 &  0.81$\pm$1.28\\
\enddata
\tablecomments{
The position differences are calculated by subtracting the values of
VLBA measurements from those of \Gaia\ EDR3, based on fitting an ring to
the SiO maser spot distributions.  Column 1, 2 and 3 list the Right
Ascension, Declination differences and the 2-dimensional position
differences.
}
\end{deluxetable*}

\section{Mira Period-Luminosity relation}
\label{sec:plr}

\begin{figure*}[ht!]
    \centering
    \includegraphics[angle=0,scale=0.44]{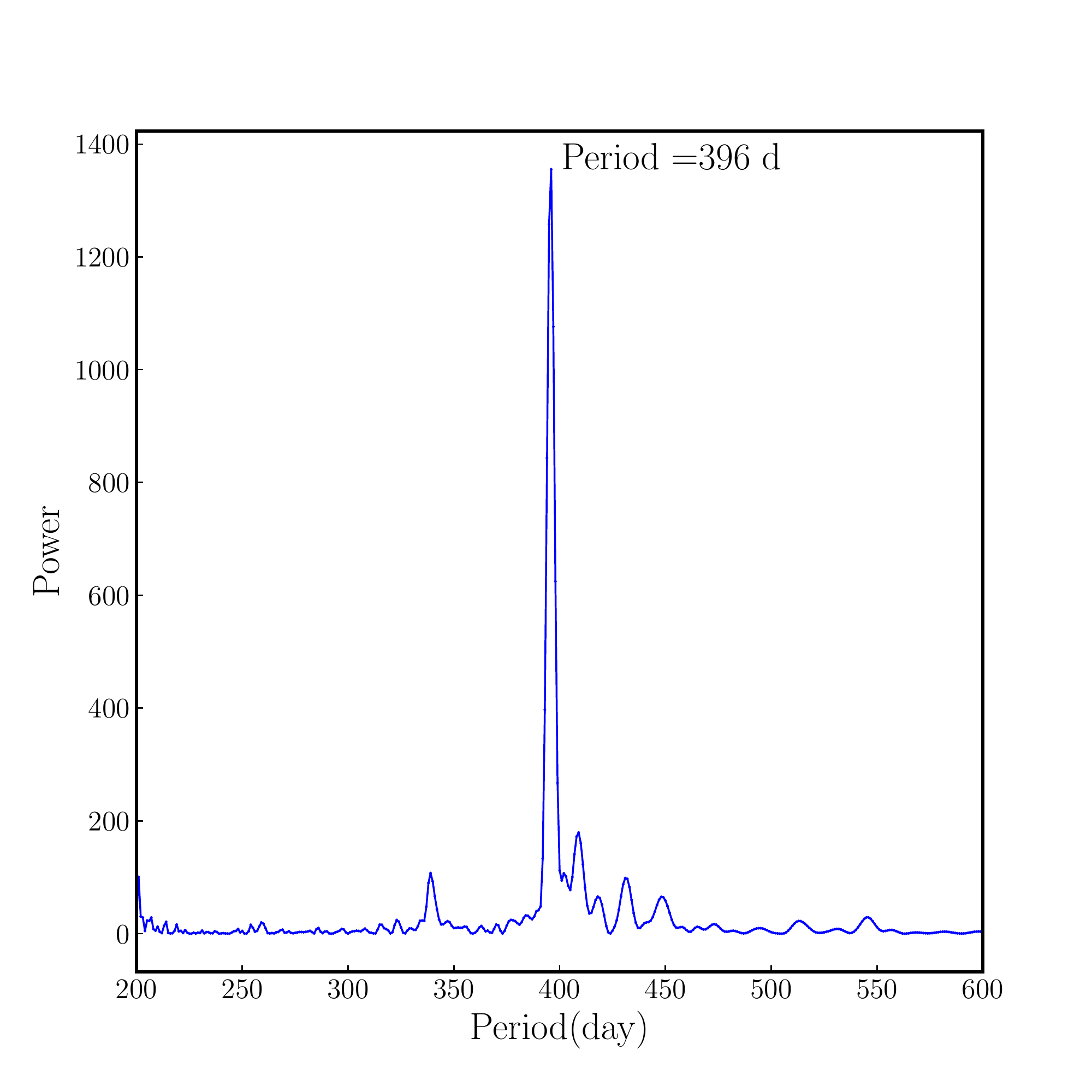}
    \includegraphics[angle=0,scale=0.44]{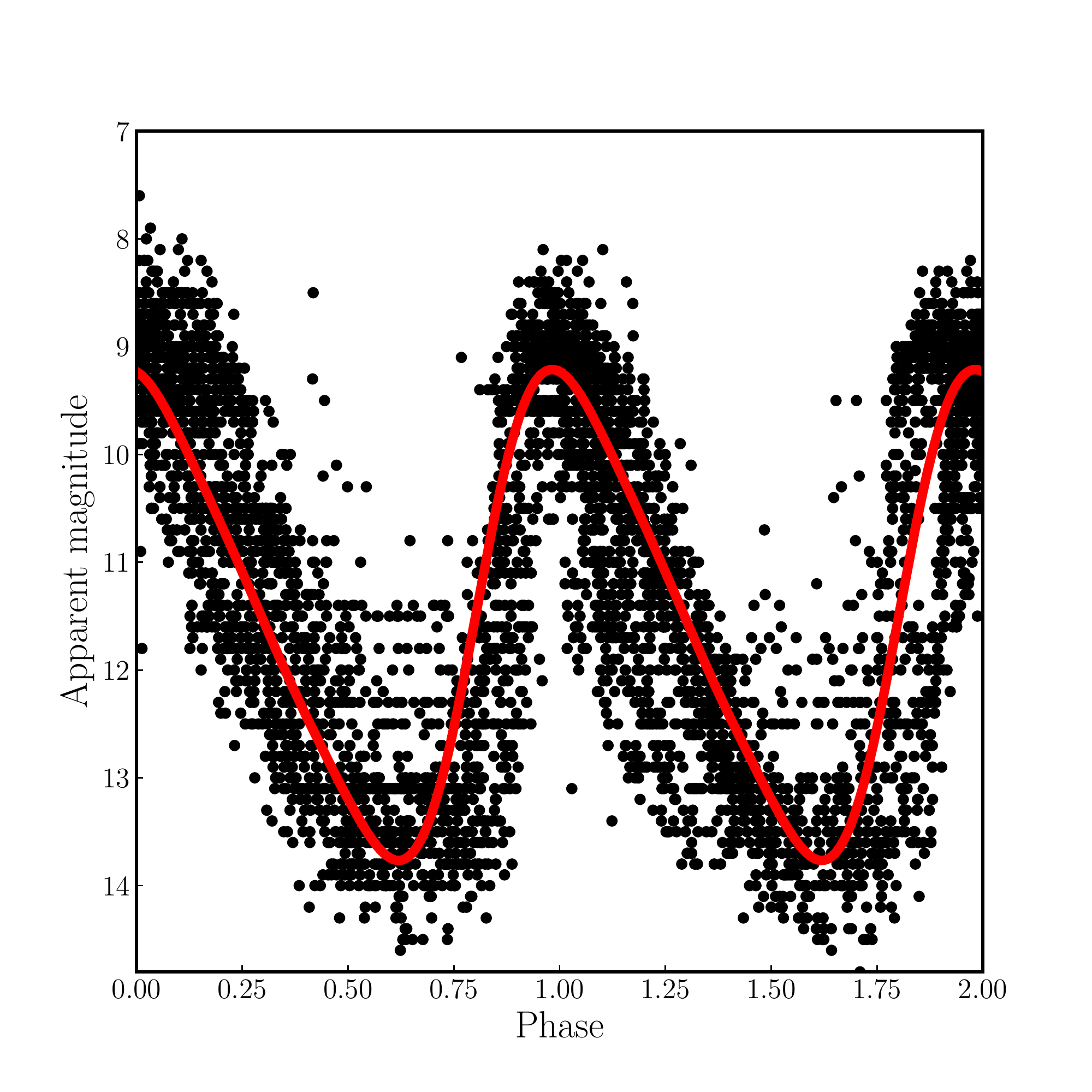}
    \caption{Period analysis result ({\it left panel}) and phase-folded light 
    curve ({\it right panel}) of the visual data from 1904 November to 2021 August 
    for RR Aql. A pulsating period of 396 days is found.
        \label{fig:period}}
\end{figure*}

As pointed out by~\citet{2000PASA...17...18W}, Miras are radial
fundamental-mode pulsators, while semi-regular variables (SRVs)
can pulsate in the fundamental mode and/or in overtones.  Pulsation
periods of SRVs can be hard to determine accurately, since as their name
indicates they often are not regular. Hence, we focus only on Miras in
this paper.
For \tsrc, in order to determine its location on the PLR diagram,  we
estimated a primary period of 396 $\pm$ 5 days (see the left
panel of Figure~\ref{fig:period}) using the visual data from 1904
November to 2021 August and the data analysis tool VStar provided
by the American Association of Variable Star Observers
(AAVSO)\footnote{\url{https://www.aavso.org}}, using the Date
Compensated Discrete Fourier Transform (DCDFT)
algorithm~\citep{1981AJ.....86..619F}, which corresponds to a maximum
likelihood sinusoidal-plus-constant regression curve-fitting.  Compared
to a DFT, the DCDFT provides a better estimation of the power spectrum
of a time series with uneven spacing. The right panel of
Figure~\ref{fig:period} shows a phase-folded visual light curve of RR
Aql using the period of 396 days.

\begin{figure}[ht!]
    \centering
    \includegraphics[angle=0,scale=0.5]{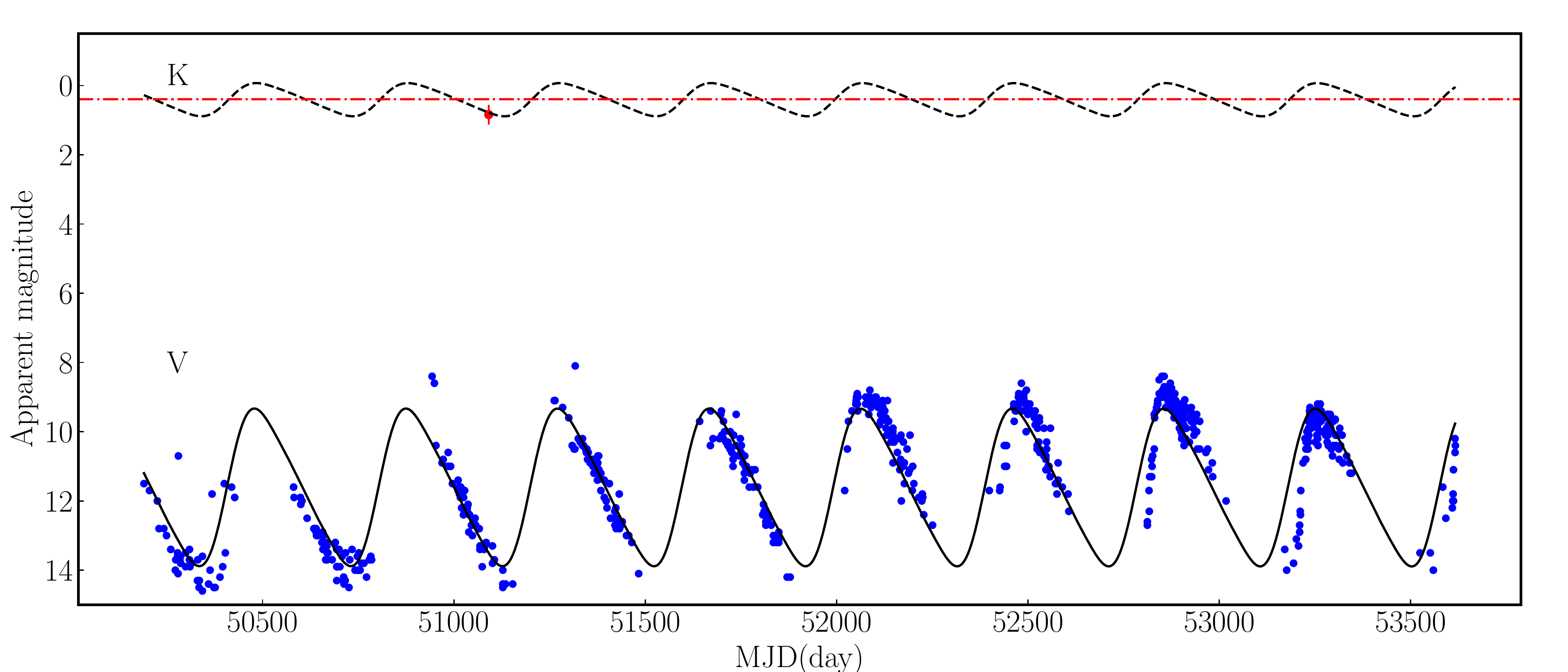}
    \caption{
    The light curve at optical $V$ ({\it lower panel}) and a model
    near-infrared $K$ band ({\it upper panel}) for RR Aql (here only the
    data from 1996 to 2005 are presented for clarity). The single-epoch
    2MASS datum is shown in red.  The solid line shows the modeled
    optical light curve estimated by VStar. The dashed line shows the
    near-infrared light curve converted from the modeled optical light
    curve. The dash-dotted line indicates the mean magnitude estimated
    by the light curve correction method for RR Aql.
    }
    \label{fig:lightcurve_fit}
\end{figure}

The apparent magnitude at an effective wavelength of $2.2$
$\mu$m for \tsrc\ (\mK = 0.84 $\pm$ 0.26 mag) comes from the Two Micron
All Sky Survey (2MASS) point-source catalog.
Since, Miras vary by upwards of 1 mag at 2$\mu$m and 2MASS measurements
are only at one epoch, this leads to extra uncertainty in the mean
magnitude owing to periodic variations. In order to better estimate a
mean apparent magnitude at near-infrared $K$ band, we adopted a method
similar to those demonstrated in~\citet{2018AJ....156..112Y} and
\citet{2021ApJS..257...23I}. This involved converting a modeled optical
light curve to a near-infrared one using an amplitude ratio
($A_{K}/A_{V}$=0.21 $\pm$ 0.09) and a phase-lag
($\Delta\Phi_{KV}$=0.10 $\pm$ 0.07) between the two
\citep{2021ApJS..257...23I}; then we added an infrared phase-adjusted
constant (--0.44 mag) to the measured magnitude. The mean magnitude for RR Aql
using the light curve correction method is 0.40 $\pm$ 0.33 mag.
Figure~\ref{fig:lightcurve_fit} shows an example of the light curve
correction to mean magnitude for RR Aql.
We estimate the uncertainty of $m^{\rm mean}_K$ by taking into account
the uncertainties from both $m_K^{\rm 2MASS}$ and the amplitude ratio
and phase-lag for light curve correction.

Table~\ref{tab:mira} lists all Miras with both VLBI and \Gaia\
parallaxes so that we could fit for PLR parameters separately for
comparison. Where available, we adopted mean magnitudes
$m_K^{\rm Kagoshima}$ of time-monitored \mK\ measurements from
the Kagoshima 1-m telescope, which used between 18 and 38 epochs per
star.  Otherwise we used $m_K^{\rm mean}$ estimated with one-epoch 2MASS
data by the light curve correction method as done for RR Aql.
Absolute magnitudes of Miras at near-infrared \K-band (\MK) were
estimated by adopting the parallax results from \Gaia\ and VLBI maser
observations, separately.
The uncertainties in the absolute magnitudes, \MK\ , takes into account
uncertainties of both \mK\ and the parallax measurements used to convert
from apparent to absolute magnitude.
All values used for Miras with both VLBI and \Gaia\ parallaxes to fit
Galactic PLRs are listed in Table~\ref{tab:mira}.

\begin{figure*}[ht!]
    \centering
    \includegraphics[angle=0,scale=0.54]{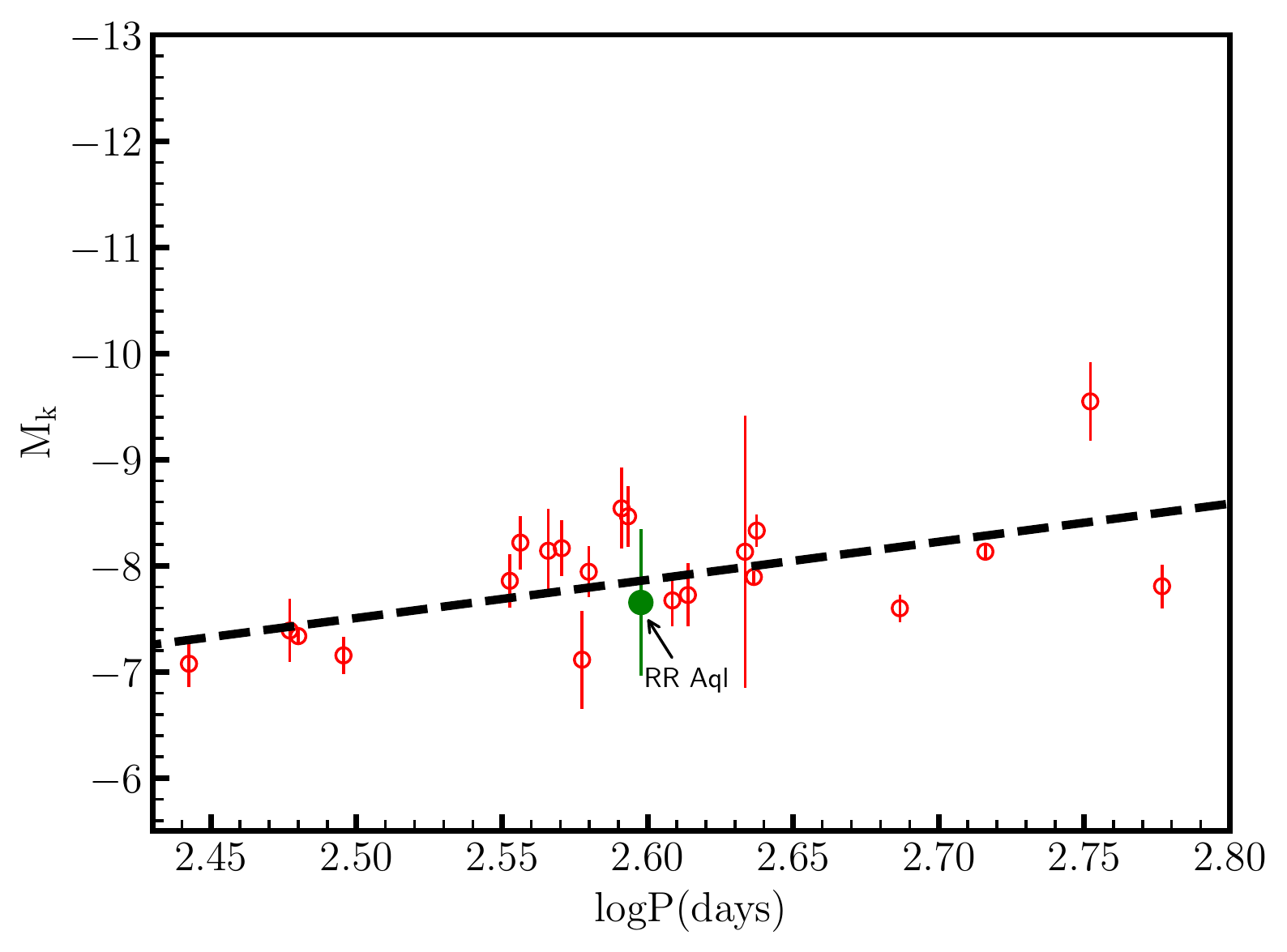}
    \includegraphics[angle=0,scale=0.54]{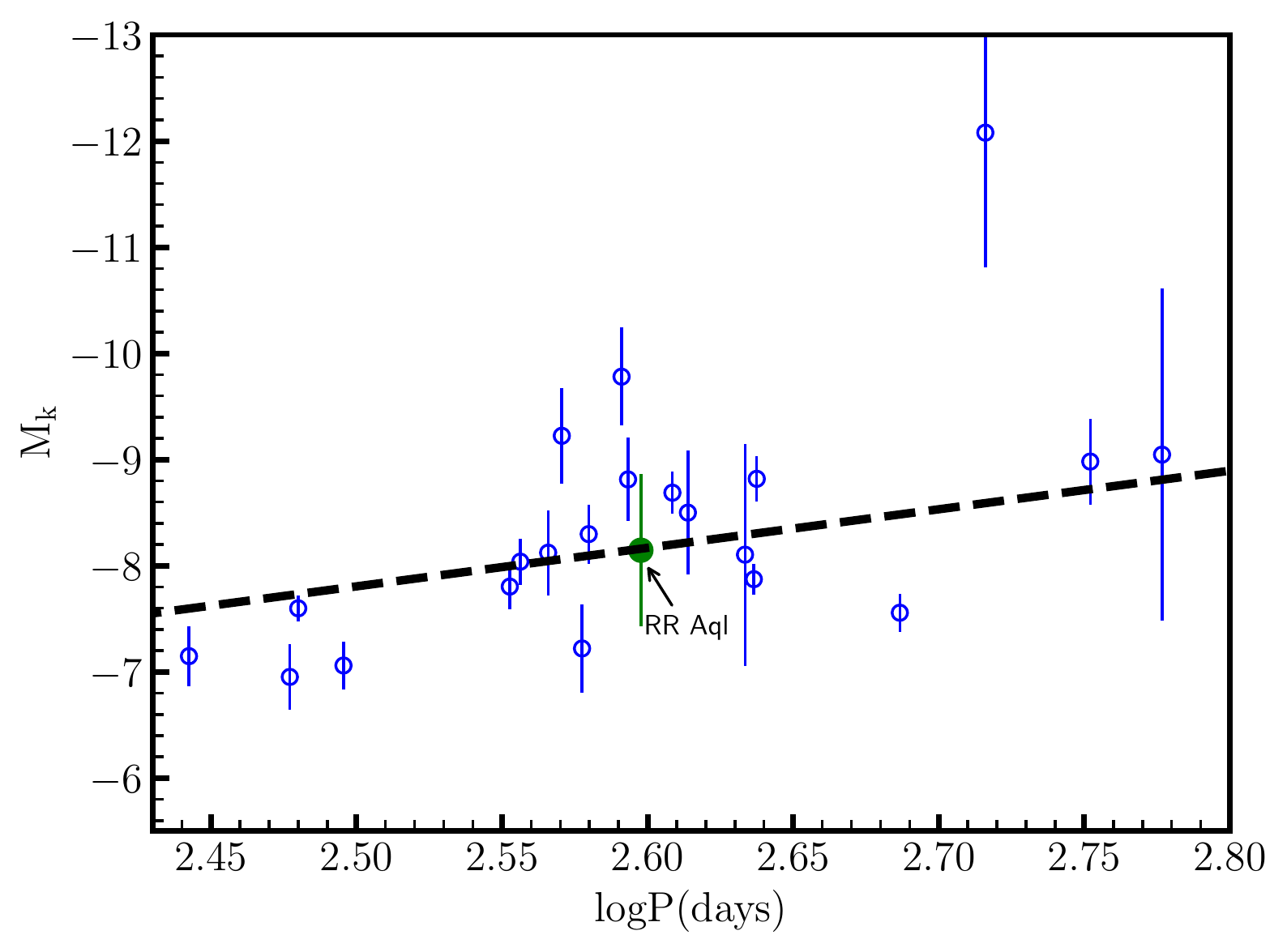}
    \caption{
    The PL(\MK) relation for Galactic Miras (dashed line)  established
    in this paper using an ``outlier tolerant'' Bayesian MCMC
    approach, fitted to data from VLBI ({\it left panel}) and \Gaia
    ({\it right panel}) parallaxes.  The location of  RR Aql on the PL
    diagram is given by a {\it filled circle} with its name.
    \label{fig:PLR}}
\end{figure*}

It is clear from Figure~\ref{fig:PLR} that, while a reasonable PLR is
evident using either VLBI or \Gaia\ parallaxes, there are some clear
outliers.  Rather than remove outliers ``by eye'', which has the
potential for significant bias, in order to fit a PLR we use a Bayesian
approach with MCMC sampling which intrinsically handles outliers
without bias. Specifically we adopt the ``conservative formulation''
of \citet{2006OUP.book......S}, for which the probability distribution
function for the uncertainty of a datum, $\sigma$, is given by $${\rm
prob}(\sigma \vert \sigma_0,I) = \sigma_0 / \sigma^2~~~,$$ where
$\sigma_0$ is a (Gaussian) uncertainty of a ``good'' datum such that
$\sigma \ge \sigma_0$.  Marginalizing over the unknown $\sigma$, leads
to a likelihood function proportional to $\ln\bigl[ {1-e^{-R^2/2} \over
R^2} \bigr]$, where $R=(D - M)/\sigma_0$ and $D$ and $M$ are the datum
and model values.  Unlike a Gaussian likelihood, this function has
Lorentzian-like tails, which assigns a significant probability to
multi-$\sigma_0$ outliers.
As prior information for our fits, we adopt a slope of --3.60 $\pm$ 0.30
and a zero point defined at logP(days) = 2.30 of --6.90 $\pm$ 1.00 mag.
Our fitted Mira PLR parameters are listed in Table~\ref{tab:PLR}.  We
note that the VLBI parallaxes yield nearly a factor of two smaller
uncertainty than the \Gaia\ parallaxes.

\startlongtable
\setlength{\tabcolsep}{2pt}
\begin{deluxetable*}{cc crc rcccrrcc c}
    \tablecaption{Miras with both VLBI and \Gaia\ parallaxes
        \label{tab:mira}
    }
    \tablehead{
        \colhead{Star}                    & \colhead{$\Pi_{\rm VLBI}$}    &
        \colhead{$\Pi_{Gaia}$}            & \colhead{$\Pi_{\rm VLBI}$-$\Pi_{Gaia}$}  & \colhead{Period}  &
        \colhead{$m_K^{\rm 2MASS}$}       & \colhead{$m_K^{\rm Kagoshima}$}   & \colhead{$m_K^{\rm mean}$}         &
        \colhead{$\sigma^{\Pi}_{M_{K}}$}  & \colhead{$M^{\rm VLBI}_{K}$}  & \colhead{$M^{\rm Gaia}_{K}$}                  & \colhead{Maser}   & \colhead{Array} & \colhead{Ref.} \\
        \colhead{    }                    & \colhead{ (mas) } &
        \colhead{ (mas)      }            & \colhead{       (mas)              }  & \colhead{ (day) } &
        \colhead{ (mag)           }       & \colhead{ (mag) }   & \colhead{ (mag) }                   &
        \colhead{ (mag)                }  & \colhead{ (mag) }   & \colhead{ (mag) }                   & \colhead{       } & \colhead{       } & \colhead{         }
    }
    \decimalcolnumbers
    \startdata
Y Lib    & 0.86$\pm$0.05   &  0.83$\pm$0.08  &   0.03  $\pm$ 0.10   & 277 &  3.08$\pm$0.29 & 3.25$\pm$0.16   &                             &  0.13    &       --7.08 $\pm$ 0.20   &   --7.15 $\pm$ 0.27 & H$_2$O & VERA & k\\ 
X Hya    & 2.07$\pm$0.05   &  2.53$\pm$0.11  &   --0.46 $\pm$ 0.12   & 300 &  1.13$\pm$0.28 &                & ~~1.03  $\pm$  0.28         &  0.05    &       --7.39 $\pm$ 0.28   &   --6.95 $\pm$ 0.30 & H$_2$O & VERA & e\\ 
R UMa    & 1.97$\pm$0.05   &  1.75$\pm$0.09  &   0.22  $\pm$ 0.10   & 302 &  1.11$\pm$0.20 & 1.19$\pm$0.02   &                             &  0.06    &       --7.34 $\pm$ 0.06   &   --7.60 $\pm$ 0.11 & H$_2$O & VERA & g\\ 
FV Boo   & 0.97$\pm$0.06   &  1.01$\pm$0.09  &   --0.04 $\pm$ 0.11   & 313 &  3.84$\pm$0.27 & 2.91$\pm$0.09  &                              &  0.13   &       --7.15 $\pm$ 0.16   &   --7.06 $\pm$ 0.21 & H$_2$O & VERA & i\\ 
R Cnc    & 3.84$\pm$0.29   &  3.94$\pm$0.18  &   --0.10 $\pm$ 0.34   & 357 & --0.71$\pm$0.17 &               &--0.78  $\pm$   0.17          &  0.16   &       --7.86 $\pm$ 0.24   &   --7.80 $\pm$ 0.20 & H$_2$O & VERA & e\\ 
S CrB    & 2.39$\pm$0.17   &  2.60$\pm$0.11  &   --0.21 $\pm$ 0.20   & 360 & --0.17$\pm$0.17 &               &--0.11  $\pm$   0.18          &  0.15   &       --8.22 $\pm$ 0.24   &   --8.04 $\pm$ 0.20 & OH     & VLBA & c\\ 
T Lep    & 3.06$\pm$0.04   &  3.09$\pm$0.10  &   --0.03 $\pm$ 0.11   & 368 & --0.27$\pm$0.35 &               &--0.57  $\pm$   0.38          &  0.03   &       --8.14 $\pm$ 0.38   &   --8.12 $\pm$ 0.39 & H$_2$O & VERA & e\\ 
S Ser    & 1.25$\pm$0.04   &  0.77$\pm$0.13  &   0.48  $\pm$ 0.14   & 372 &  1.68$\pm$0.18 &                 &~~1.35  $\pm$   0.24         &  0.07    &       --8.17 $\pm$ 0.25   &   --9.22 $\pm$ 0.44 & H$_2$O & VERA & e\\ 
R Peg    & 2.76$\pm$0.28   &  2.63$\pm$0.12  &   0.13  $\pm$ 0.30   & 378 &  0.38$\pm$0.36 &                 &~~0.68  $\pm$   0.39         &  0.22    &       --7.12 $\pm$ 0.45   &   --7.22 $\pm$ 0.40 & H$_2$O & VERA & e\\ 
R Hya    & 7.93$\pm$0.18   &  6.74$\pm$0.46  &   1.19  $\pm$ 0.50   & 380 & --2.66$\pm$0.19 &                &--2.44  $\pm$   0.22          &  0.05   &       --7.94 $\pm$ 0.23   &   --8.30 $\pm$ 0.27 & H$_2$O & VERA & e\\ 
R Aqr    & 4.59$\pm$0.24   &  2.59$\pm$0.33  &   2.00  $\pm$ 0.41   & 390 & --1.60$\pm$0.33 &                &--1.85  $\pm$   0.35         &  0.11    &       --8.54 $\pm$ 0.37   &   --9.78 $\pm$ 0.45 & SiO    & VERA & e\\ 
W Leo    & 1.03$\pm$0.02   &  0.88$\pm$0.11  &   0.15  $\pm$ 0.11   & 392 &  1.80$\pm$0.23 &                 &~~1.47  $\pm$   0.27        &  0.04     &       --8.47 $\pm$ 0.27   &   --8.81 $\pm$ 0.38 & H$_2$O & VERA & e\\ 
RR Aql   & 2.45$\pm$0.08   &  1.95$\pm$0.11  &   0.50  $\pm$ 0.14   & 396 &  0.84$\pm$0.26 &                 &~~0.40  $\pm$   0.33        &  0.07     &       --7.65 $\pm$ 0.34   &   --8.15 $\pm$ 0.35 & H$_2$O & VLBA & d\\ 
U Her    & 3.76$\pm$0.27   &  2.36$\pm$0.08  &   1.40  $\pm$ 0.28   & 406 & --0.63$\pm$0.16 &                &--0.55  $\pm$   0.17         &  0.16    &       --7.67 $\pm$ 0.23   &   --8.69 $\pm$ 0.18 & OH     & VLBA & c\\ 
SY Scl   & 0.75$\pm$0.03   &  0.52$\pm$0.12  &   0.23  $\pm$ 0.13   & 411 &  2.65$\pm$0.24 &                 &~~2.90  $\pm$   0.27        &  0.09     &       --7.72 $\pm$ 0.28   &   --8.50 $\pm$ 0.57 & H$_2$O & VERA & e\\ 
R Cas    & 5.67$\pm$1.95   &  5.74$\pm$0.20  &   --0.07 $\pm$ 1.96   & 430 & --1.40$\pm$1.00 &               &--1.90  $\pm$   1.03          &  0.75   &       --8.13 $\pm$ 1.27   &   --8.10 $\pm$ 1.03 & OH     & VLBA & a\\ 
AP Lyn   & 2.00$\pm$0.04   &  2.02$\pm$0.12  &   --0.02 $\pm$ 0.13   & 433 &  0.98$\pm$0.20 & 0.60$\pm$0.01  &                              &  0.04   &       --7.90 $\pm$ 0.05   &   --7.87 $\pm$ 0.13 & H$_2$O & VERA & h\\ 
U Lyn    & 1.27$\pm$0.06   &  1.01$\pm$0.08  &   0.26  $\pm$ 0.10   & 434 &  1.53$\pm$0.23 & 1.15$\pm$0.09   &                             &  0.10    &       --8.33 $\pm$ 0.14   &   --8.82 $\pm$ 0.20 & H$_2$O & VERA & f\\ 
BX Cam   & 1.73$\pm$0.03   &  1.76$\pm$0.10  &   --0.03 $\pm$ 0.11   & 486 &  0.91$\pm$0.19 & 1.21$\pm$0.11  &                              &  0.04   &       --7.60 $\pm$ 0.12   &   --7.55 $\pm$ 0.17 & H$_2$O & VERA & h\\ 
V837 Her & 1.09$\pm$0.02   &  0.18$\pm$0.10  &   0.91  $\pm$ 0.10   & 520 &  2.06$\pm$0.27 & 1.68$\pm$0.03   &                             &  0.04    &       --8.13 $\pm$ 0.05   &   --12.08$\pm$ 1.25 & H$_2$O & VERA & h\\ 
UX Cyg   & 0.54$\pm$0.06   &  0.70$\pm$0.09  &   --0.16 $\pm$ 0.11   & 565 &  1.40$\pm$0.19 &                &~~1.79  $\pm$   0.26          &  0.24   &       --9.55 $\pm$ 0.35   &   --8.98 $\pm$ 0.39 & H$_2$O & VLBA & b\\ 
OZ Gem   & 0.81$\pm$0.04   &  0.45$\pm$0.33  &   0.35  $\pm$ 0.33   & 598 &  3.00$\pm$0.35 & 2.65$\pm$0.16   &                             &  0.11    &       --7.81 $\pm$ 0.19   &   --9.05 $\pm$ 1.55 & H$_2$O & VERA & j\\
    \enddata
    \tablecomments{
        Column 1 lists the names of Miras, columns 2 and 3 list VLBI and
        \Gaia\ parallaxes, respectively, and column 4 lists VLBI and
        \Gaia\ parallax differences.
        Column 5 lists the periods of magnitude variation at optical
        wavelength from AAVSO, except for V837 Her and AP Lyn, for which
        periods are from~\citet{2020PASJ...72...59C}.  Apparent
        magnitudes at infrared $K$-band are listed in column 6, 7 and 8:
        $m_K^{\rm 2MASS}$ values are from the 2MASS point-source
        catalog~\citep{2006AJ....131.1163S}, $m_K^{\rm Kagoshima}$ are
        from the time-monitored observations with the Kagoshima
        University 1-m telescope and $m_K^{\rm mean}$ are the
        mean apparent magnitudes estimated using the light curve
        correction method.
        Adopted to calculate the absolute magnitude \MK\ (listed in
        column 10 and 11) is $m_K^{\rm Kagoshima}$ if available,
        otherwise $m_K^{\rm mean}$. The uncertainties of \MK\ owing to
        VLBI parallax uncertainties are listed in column 9. The maser
        species are listed in column 12. The last two columns list the
        VLBI array for parallax measurements and references,
        respectively.
         The \Gaia\ parallaxes are from~\citet{{2021A+A...649A...1G}}.
         The VLBI parallaxes are from
         (a)~\citet{{2003A+A...407..213V}},
         (b)~\citet{2005ApJ...627L..49K},
         (c)~\citet{{2007A+A...472..547V}},
         (d)~this paper,
         (e)~\citet{2020PASJ...72...50V},
         (f)~\citet{2016PASJ...68...71K},
         (g)~\citet{2016PASJ...68...78N},
         (h)~\citet{2020PASJ...72...59C},
         (i)~\citet{2016PASJ...68...75K},
         (j)~\citet{2020PASJ...72...57U},
         (k)~\citet{2019PASJ...71...92C},
         (l)~\citet{2014PASJ...66..107K}.
        }
\end{deluxetable*}

As shown by \citet{2021ApJ...919...99I} and \citet{2021ApJS..257...23I},
a sample of O-rich Miras with a large period range clearly shows a
non-linear PLR, while a sample restricted to log $P$ $\approx$ 2.0 - 2.8
can be well described by a linear PLR. To compare our  Galactic PLR with
an even smaller range of log $P$ from $\approx$ 2.4 - 2.8 to the
existing PLRs in literature, we list only the latest determined linear
PLR of O-rich Miras measured at near-infrared $K$-band for different
galaxies in Table~\ref{tab:PLR}.
\citet{2017AJ....154..149Y} determined a linear PLR based on about 160
O-rich Miras with period $\textless$ 400 days at near-infrared
wavelengths, and we adjusted the zero-point using the updated distance
modulus of 18.477 $\pm$ 0.026 mag for LMC from
\cite{2019Natur.567..200P}.  We took account of uncertainties from both
the photometric zero-point ($\sim$ 0.02 mag) and distance modulus
($\sim$ 0.03 mag), which were not included
in~\citet{2017AJ....154..149Y}. 
The linear PLR for M~33 \citep{2018AJ....156..112Y} is from
near-infrared light curves of 1169 O-rich Miras with
periods $\textless$ 400 days assuming a slope of --3.77, which is the
same as the LMC value from~\citet{2017AJ....154..149Y}.
In addition, assuming a distance modulus of 18.477 $\pm$ 0.026 mag from
\cite{2019Natur.567..200P}, \citet{2021ApJS..257...23I} determined the
PLR of Miras in LMC at near-infrared $K$-band using a sample of 29
O-rich Miras.

Since we rely on a strong prior for the slope of the PLR, we now only
compare zero-points for the LMC and M~33 and with Milky Way values based
on VLBI and Gaia calibrations. It is clear that the PLR zero-points of LMC and M~33 derived by
\cite{2017AJ....154..149Y,2018AJ....156..112Y} differ by only $0.05$ mag
and are consistent with each other. The zero-points for the Milky Way
based on VLBI and Gaia parallaxes are statistically consistent with each
other, with the VLBI value being more accurate.
Compared with our Galactic PLR using VLBI parallaxes, the LMC zero-point
derived by~\citet{2017AJ....154..149Y} differs by 0.13 mag. The PLR
zero-point derived by \citet{2021ApJS..257...23I} has a larger
uncertainty and differs by 0.12 mag.  All these zero-point differences
are consistent within their joint uncertainties.

We derive ``parallaxes'' (from PLR-based distances) of 2.05 $\pm$ 0.31
mas and 2.45 $\pm$ 0.41 mas for RR Aql, using the LMC $K$-band PLRs of
\citet{2017AJ....154..149Y} and \citet{2021ApJS..257...23I},
respectively.  While the latter PLR-based parallax is close to our VLBI
parallax of 2.44 $\pm$ 0.07 mas, both are consistent within their
uncertainties.

The scatter of PLR post-fit residuals listed in Table~\ref{tab:PLR} for
Galactic Miras is larger than for the LMC or M~33. Two likely reasons
include 1) the Miras in other galaxies are at nearly the same distance,
which removes distance uncertainty as a source of scatter, and 2) more
than half of the Galactic Miras listed in Table~\ref{tab:mira} have only
a single epoch of 2MASS magnitudes and these were measured with
saturated detectors. 
The 2MASS apparent magnitude for these Miras at near-infrared
\K-band are brighter than $4^{th}$ mag, which is the saturation limit for the 51
ms exposures, leading to larger magnitude uncertainty. Fluxes for these
Miras were estimated by template fitting to the unsaturated scattered 
light in the wings of the saturated star
image~\citep{2006AJ....131.1163S}.
%
An examination of Table~\ref{tab:mira}, suggests a typical uncertainty
of the mean apparent 2MASS magnitudes is $\approx0.3$ mag.    Removing a
scatter of this magnitude from the total scatter of 0.45 mag for the
VLBI calibrated PLR, we find a residual scatter of 0.34 mag (ie,
$\sqrt{0.45^2-0.3^2}$).  This residual scatter is currently unaccounted
for, but could come from slightly underestimated parallax uncertainties
and/or intrinsic brightness variations over different stellar cycles.

\begin{deluxetable*}{cc c c c c}
    \tablecaption{Coefficients of PLR of Miras at  near-infrared $K$-band in different galaxies}
    \label{tab:PLR}
    \tablewidth{0pt}
    \tablehead{
    \colhead{Galaxies}  & \colhead{Slope ($a$)} & \colhead{Zero-point} ($b$) & \colhead{Scatter} & \colhead{$N$} & \colhead{References} \\
        \colhead{}                & \colhead{}            & \colhead{(mag)}            & \colhead{(mag)}   & \colhead{}    & \colhead{}
}
    \decimalcolnumbers
    \startdata
    LMC         & --3.77 $\pm$ 0.07   &  --6.92 $\pm$ 0.04 & 0.12 & 158  & \citet{2017AJ....154..149Y} \\
    M~33        & [--3.77]            &  --6.97 $\pm$ 0.01 & 0.21 & 1169 & \citet{2018AJ....156..112Y} \\
\\
    LMC         & --3.30 $\pm$ 0.46   &  --6.67 $\pm$ 0.06 & 0.27 & 29   & \citet{2021ApJS..257...23I} \\
\\
     Milky Way  & --3.59 $\pm$ 0.29   &  --6.79 $\pm$ 0.15 & 0.45 & 22   & This paper using VLBI parallaxes \\
     Milky Way  & --3.63 $\pm$ 0.30   &  --7.08 $\pm$ 0.29 & 0.94 & 22   & This paper using \Gaia\ parallaxes \\
    \enddata
    \tablecomments{
    PLRs are defined as $M= a ({\rm logP(days)} - 2.30) + b$ for Miras
    at near-infrared $K$-band, matching those of
    \citet{2017AJ....154..149Y, 2018AJ....156..112Y} and
    \citet{2021ApJS..257...23I}.  Columns 2 and 3 give the PLR slopes
    and zero-points.  The Milky Way fits used a strong prior of --3.60
    $\pm$ 0.30 for the slope and a very weak prior of --6.90 $\pm$ 1.00
    mag for the zero-points.  Columns 4 and 5 list the standard
    deviation of the  post-fit residuals in magnitudes and the number of
    Miras fitted.
    The results for the LMC by \citet{2017AJ....154..149Y} were based on
    Miras with periods $\textless~400$ days, assuming distance modulus
    of 18.493 $\pm$ 0.048  mag \citep{2013Natur.495...76P}; we have
    adjusted the zero-point and its uncertainty to an updated distance
    modulus of 18.477 $\pm$ 0.026 mag \citep{2019Natur.567..200P}.
    The slope of the PLR for M~33 by \citet{2018AJ....156..112Y} was
    fixed at --3.77 to match the LMC slope determined by
    \citet{2017AJ....154..149Y}.
    The PLR for the LMC by \citet{2021ApJS..257...23I} assumed a
    distance modulus of 18.477 $\pm$ 0.026 mag
    \citep{2019Natur.567..200P}; the zero-point uncertainty includes
    both the statistical and systematic errors.
    }
\end{deluxetable*}

\section{Conclusions}
\label{sec:sum}

We have measured the trigonometric parallax and proper motion of the
Mira variable RR Aql using VLBA observations of its water and SiO
masers.  Our parallax is 2.44 $\pm$ 0.07 mas, which differs
significantly from the \Gaia\ EDR3 parallax of 1.95 $\pm$ 0.11 mas.  We
fitted the positions of SiO masers to a circle and obtained the absolute
position of the central star.  Comparing the absolute positions of RR
Aql from VLBA and \Gaia, we find agreement within their  joint
uncertainties.

%
Using 22 Miras, including \tsrc, which have both VLBI and \Gaia\
parallaxes, we have derived separately PLRs of Miras at near-infrared K
band.  The post-fit residuals using VLBI parallaxes are about half as
large as those using \Gaia\ parallaxes, suggesting that VLBI parallaxes
are more accurate than \Gaia\ parallaxes for Miras.   Thus, VLBI
astrometry should be effective to validate future \Gaia\ results for
Miras.

%

Compared with the PLRs for Miras in other galaxies, the Galactic PLR
based on the VLBI and \Gaia\ parallaxes are statistically consistent
with PLRs in LMC and M~33 at near-infrared \K-band.
Since Miras vary by about 1 mag at infrared $K$-band, absolute
magnitudes of Miras are uncertain by about $\pm~0.4$ mag when photometry
from only a single epoch is used. This often dominates over parallax
uncertainty when estimating absolute magnitudes.  Thus, priority should
be given to monitoring surveys of Miras at infrared $K$-band in order to
obtain accurate measurements of mean \mK.  Lacking multi-epoch
magnitudes, combining optical light curves with infrared magnitudes can
help to reduce the uncertainty due to the periodic variation.

%
In the era of space infrared telescopes, e.g., JWST, one could
obtain extremely accurate apparent magnitudes for Miras.  In addition,
future astrometric results of \Gaia, which could be validated by VLBI
observations, are expected to greatly increase the number of Miras with
accurate parallaxes. More reliable Galactic PLRs of Miras will be
available, leading to Miras being widely used as distance indicators.

\begin{acknowledgements}

This work was partly supported by the National Science Foundation of
China under grant U1831136 and U2031212, and the Key Laboratory
for Radio Astronomy, Chinese Academy of Sciences.

\end{acknowledgements}

\facilities{VLBA}

\software{DiFX~\citep{2007PASP..119..318D},
AIPS~\citep{1996ASPC..101...37V}}

\clearpage
\bibliography{ref}
\bibliographystyle{aasjournal}

\end{document}